\documentclass[12pt]{article}

\usepackage{amssymb,amsmath,amsfonts,eurosym,geometry,ulem,graphicx,caption,color,setspace,sectsty,comment,footmisc,caption,pdflscape,array,subcaption}
\usepackage{tabularx,multirow}
\usepackage{authblk}
\usepackage{booktabs}
\usepackage[authoryear,round,numbers]{natbib}
\usepackage[export]{adjustbox}
\usepackage[colorlinks = true,
            linkcolor = blue,
            urlcolor  = blue,
            citecolor = blue,
            anchorcolor = blue]{hyperref}
\usepackage[ruled,vlined,linesnumbered]{algorithm2e}

\newtheorem{hyp}{Hypothesis}

\newcolumntype{L}[1]{>{\raggedright\let\newline\\arraybackslash\hspace{0pt}}m{#1}}
\newcolumntype{C}[1]{>{\centering\let\newline\\arraybackslash\hspace{0pt}}m{#1}}
\newcolumntype{R}[1]{>{\raggedleft\let\newline\\arraybackslash\hspace{0pt}}m{#1}}

\newcommand*\samethanks[1][\value{footnote}]{\footnotemark[#1]}
\begin{document}

\title{\vspace*{-2\baselineskip}End-to-End Risk Budgeting Portfolio Optimization with Neural Networks}
\author[1]{A. Sinem Uysal\thanks{These two authors contributed equally}\thanks{\url{auysal@princeton.edu}}}
\author[1]{Xiaoyue Li\samethanks[1]\thanks{\url{xiaoyuel@princeton.edu}}}
\author[1]{John M. Mulvey\thanks{The Bendheim Center for Finance, Center for Statistics and Machine Learning, Princeton University}}

\affil[1]{Department of Operations Research and Financial Engineering, Princeton University}

\date{\today \vspace{-1.5em}}

\maketitle

\begin{abstract}
\noindent Portfolio optimization has been a central problem in finance, often approached with two steps: calibrating the parameters and then solving an optimization problem. Yet, the two-step procedure sometimes encounter the ``error maximization'' problem where inaccuracy in parameter estimation translates to unwise allocation decisions. In this paper, we combine the prediction and optimization tasks in a single feed-forward neural network and implement an end-to-end approach, where we learn the portfolio allocation directly from the input features. Two end-to-end portfolio constructions are included: a model-free network and a model-based network. The model-free approach is seen as a black-box, whereas in the model-based approach, we learn the optimal risk contribution on the assets and solve the allocation with an implicit optimization layer embedded in the neural network. The model-based end-to-end framework provides robust performance in the out-of-sample (2017-2021) tests when maximizing Sharpe ratio is used as the training objective function, achieving a Sharpe ratio of 1.16 when nominal risk parity yields 0.79 and equal-weight fix-mix yields 0.83. Noticing that risk-based portfolios can be sensitive to the underlying asset universe, we develop an asset selection mechanism embedded in the neural network with stochastic gates, in order to prevent the portfolio being hurt by the low-volatility assets with low returns. The gated end-to-end with filter outperforms the nominal risk-parity benchmarks with naive filtering mechanism, boosting the Sharpe ratio of the out-of-sample period (2017-2021) to 1.24 in the market data.\\
\vspace{-1em}

\end{abstract}
\setcounter{page}{0}
\thispagestyle{empty}

\pagebreak \newpage
\doublespacing
\section{Introduction}
Machine learning and deep learning algorithms have become  common methodological approaches implemented by researchers in various domains due to advances in big data and computational tools. Similarly, quantitative finance has utilized  these methods to bring new approaches to various domain-specific problems in finance. Predictive modeling tasks are one of the main areas where machine learning methods and deep learning have been  employed extensively. These methods not only help to improve model performances but also allow to explore different datasets such as high dimensional or alternative data. However, in many problems prediction is not the ultimate goal, but instead a part of a larger process. Similarly, financial decision making systems can be divided into prediction and optimization tasks, where the learning methods are usually applied to address the former one.\\
\\
Portfolio optimization is a famous problem in finance which is formally started with Markowitz portfolio theory (\cite{Markowitz1952}), and it can be separated into two parts: prediction and optimization. Some of the common prediction tasks are: asset parameter predictions for the  optimization problem (e.g. factor models, equity return prediction, covariance matrix structuring), regime prediction for regime-switching portfolios, etc. It has been shown that these approaches can improve prediction performance compared to traditional models. However, in portfolio optimization, asset parameter prediction is not the end goal, but rather an intermediate step for asset allocation decision task. General approach to tackle this problem is to model in two stages: (1) generate model parameters with predictive models, (2) incorporate  prediction results in decision making for portfolio construction via an optimization problem. If the predictive model has good performance, the two stage approach is beneficial. When the predictive model accuracy is not good, which is usually the case in asset return prediction, the two stage models can lead to sub-optimal decisions for the main task. This problem occurs due to training for different goals in each stages. Prediction models are trained to minimize prediction errors, and optimization problems are optimized usually for decision loss functions. Therefore, the two stage models can lead to error accumulation which result in suboptimal decisions.\\
\\
In this paper, our goal is to tackle the portfolio optimization problem in one stage with an end-to-end learning approach. Here we combine the prediction and optimization tasks in a single fully connected neural network and generate portfolio decisions as outputs. 
We do not train the values of the relevant parameters directly. Instead, we leave it to the neural network to choose the optimal values for the estimation. After the hidden layers, we present two approaches for end-to-end portfolio construction: model-free and model-based networks. Model-free networks are trained for different cost functions based on portfolio metrics, and model-based networks employ a specific portfolio optimization model in the network via differentiable optimization layers. The former approach is seen as black-box or model-free models which is less interpretable. On the other hand, model-based networks employ a specific rule for portfolio decision which makes them more interpretable in terms of model approximation. The second layer in model-based networks are different from hidden layers, which represents the optimization problem of our choice. In particular, this layer is constructed as a parametrized optimization problem where parameters depend on the previous layer, and the cost function of the network is a  portfolio-based performance metric. Furthermore, we present a novel way to introduce an asset selection property into the risk budgeting portfolios. Embedded stochastic gates allow filtering the undesired assets from the portfolio, which helps to boosts the portfolio performance and beats the nominal benchmarks in the market data.\\
\\
There are exciting developments on end-to-end learning approaches in order to integrate prediction and decision making tasks. In this paper, we follow \cite{Donti2017} and \cite{Agrawal2019}, and propose an end-to-end portfolio construction framework. Here we choose risk budgeting portfolio optimization (\cite{Bruder2012}) as the policy rule in the end-to-end learning system, and implement it as an differentiable optimization layer in the neural network. Risk budgeting portfolio technique is a family of policy rules for constructing risk-based portfolios, some of which are risk-based diversification without considering expected asset return. Risk budgeting portfolios enjoy several desired properties that the classical Markowitz does not have, and is immune to the difficulty of return estimation. Equal risk contribution or risk parity is a special version of the risk budgeting portfolios where all risk budgets are set equal (\cite{Maillard2010}). It has drawn increasing attention due to its robustness under various market environments and its low sensitivity to parameter estimations (\cite{Ardia2017}). According to Neuberger Berman Asset Management (\cite{Kaya2020}) there is about \$120 billion asset under management based on risk parity strategies (\cite{Fabozzi2021}). In addition, this method may be employed with different portfolio optimization techniques according to investors' preference. In this paper, we present a framework that integrates prediction and optimization steps in portfolio construction, and that can also be extended to many other quantitative finance problems which involves prediction and optimization tasks.\\
\\
 The rest of the paper is organized as follows. Section \ref{sec:litreview} highlights the previous work on end-to-end learning and presents different approaches. Section \ref{sec:model} explains the end-to-end portfolio model method with neural networks and optimization layers. Section \ref{sec:sim-study} shows results from the simulation study, and Section \ref{sec:market-data} presents computational results of the extensive experiments on the real market data. Section \ref{sec:conclusion} concludes with future work directions.

\section{Literature Review}\label{sec:litreview}
There is a growing body of literature on data-driven decision making contributed by various research fields. Researchers have become interested in connecting prediction and decision making stages in the problem at hand. In the past five years, various approaches are proposed to integrate prediction and optimization tasks. Herein, we will explain the methods whose main goal is to connect these two stages in the decision making problems.\\
\\
\cite{Donti2017} present end-to-end training approach within stochastic optimization models where learning is performed based on task loss instead of generic loss function. Prediction models are part of a larger process and probabilistic machine learning models are trained to capture task-based objective. They report improved experimental results in inventory stock problem, energy scheduling and energy storage tasks in comparison to traditional maximum likelihood maximization method and a black-box neural network approach.
\cite{wilder2019} consider combinatorial optimization problems in the data driven decision making pipeline where they compare performances of two-stage and decision-focused models. In the latter approach, machine learning models are trained based on their performance on combinatorial optimization tasks. They report that decision-focused models outperform two-stage ones, especially in low-signal environments.\\
\\
\cite{Elmachtoub2017} introduce a framework called "Smart Predict then Optimize (SPO)" to leverage optimization problem structure to design better prediction models. They replace the prediction based loss function with the decision error by using a convex surrogate loss function which is applicable to linear models. Later \cite{Elmachtoub2020} consider the use of decision trees for decision-making problems under the predict-then-optimize framework. In the recent work \cite{balghatibounds}, they provide generalization  bounds for the predict then optimize framework.\\
\\
An interesting line of research focuses on embedding the optimization problem in the neural networks which is referred as optimization or implicit layers. This approach allows to integrate the parametrized optimization problem as an individual layer in an end-to-end trainable neural network whose parameters are learned through the propagation.  \cite{Amos2017} propose a network architecture, OptNet, which integrates quadratic programs as an individual layer whose parameters can depend on previous layer in any differentiable way. The layers are learned by taking gradients of the loss function with respect to the parameters. Karush-Kuhn-Tucker conditions are used to derive gradients of the optimization layer in the back-propagation step. \cite{Agrawal2019} generalize this work and introduce convex optimization layers. They provide a software package called  \texttt{CvxpyLayer} based on the convex optimization library \texttt{cvxpy} and includes implementations for the famous deep learning libraries, \texttt{PyTorch} and \texttt{TensorFlow}. Here differentiable convex optimization layers are constructed by following disciplined parametrized programming rules and tackle all convex optimization problems that can be written in this framework. They present applications in linear machine learning models and in stochastic control. Aforementioned optimization layers are valid for convex optimization problems. However, \cite{Gould2019} discuss the use of non-convex optimization problems as implicit layers.\\
\\
\cite{Bertsimas2020} connect approaches from machine learning and operations research fields, and propose a different framework from the previous studies to integrate these two tasks. They suggest a method called "Prescriptive Analytics" that incorporates auxiliary information in data-driven decision making system. They emphasize that the main focus of supervised learning systems is to provide optimal predictions, not optimal decision making under uncertainty.
They introduce a link function to generate data-driven prescriptive predictions, and provide specific constructions for a great variety of supervised learning models with theoretical guarantees. A real-world example from inventory management problem demonstrates the benefits of their proposed framework. \\
\\
There has been some work in finance with similar objectives, integrating prediction and decision steps in a data-driven decision making system within a neural network. The earliest work by \cite{Bengio1996} emphasizes the importance of training models in order to optimize financial criteria of interest, especially in noisy time series data. He designs two modules in a fully connected network with a single hidden layer where the first module produces stock return predictions and the second module is a trading module based on a prior knowledge. Results from portfolio selection on 35 Canadian stocks shows benefits of the joint training of the modules. 
Recently a similar approach appears in \cite{Zohren2020} where different neural networks are trained to optimize portfolio Sharpe ratio. Out of various network architectures, they found Long Short-term Memory model to be the best performing one. Asset price and returns with current and past 50 day values are fed as the features in the model and they report overfitting with fully connected network  due to large number of parameters.  \cite{Butler2021} integrates prediction and optimization framework in the mean-variance portfolio problem. They provide closed-form solutions under certain constraints, and use neural networks with differentiable optimization layers to find the solution where closed-form solution doesn't exist. The real-world empirical tests show gains in the performance over two-step models, where the traditional approach was performing linear regression then portfolio optimization.\\
\\
A similar end-to-end learning approach appears in the reinforcement learning field where raw observations are directly mapped to actions with model-based and model-free methods. The goal is to find a policy that maximizes the cumulative reward achieved.  \cite{Amos2018} propose using model predictive control as a differentiable policy class in reinforcement learning. They found this approach much less computational and memory intensive compare to the traditional approach. A recent trading model application of reinforcement learning can be found in \cite{Zhang2020-RLtrading}. However, the main goal in this area is not to connect two-stage processes. Therefore, we omit the review of this literature. 
\subsection{Our Approach}
In this paper, we adopt an end-to-end framework to learn the investment strategy based on risk budgeting portfolio optimization model. We follow end-to-end decision making pipeline approach by \cite{Donti2017} and \cite{Agrawal2019} and construct a neural network with an implicit layer to embed the risk budgeting portfolio optimization problem.  Both model-free and model-based structures are tested. In the model-free strategy, a feed-forward neural network directly learns from the features and outputs the allocation decision. In the model-based strategy, the neural network first learns the risk contributions of each asset from the features, based on which allocations are made. Risk-based portfolios are found to be more robust to various market environment as well as to the errors in parameter estimation, but is known to be sensitive to the underlying asset universe. With a risk-budgeting model-based approach, we aim to inherit the robustness of risk-based portfolios, in the meanwhile shifting away from the undesired assets by dynamically allocating low risk budgets. Finally, we introduce a novel asset selection feature into the end-to-end system with stochastic gates to construct sparse portfolios that are robust to the underlying asset universe. Adding the filtering property boosts the performance in the market data and helps to protect the risk budgeting portfolio against unprofitable low volatility assets.

\section{Methodology}\label{sec:model}
A widely adopted approach in portfolio optimization is to estimate the relevant parameters, such as expected returns and covariance matrix, by predictive models and make allocation decisions based on the parameters to optimize an objective function. The two-step procedure has been employed by practitioners and researchers for a long time. Yet, it faces two critics by nature. One critic of such methods is that it heavily depends on intermediate estimations, and some portfolio frameworks are unstable with respect to errors in the intermediate estimations. Such indirect optimization may bring inconsistency of goals and lead to suboptimal decisions. In the estimation step, data are summarized into a few key features, leaving out other potentially helpful information. Second, literature shows it is generally hard to provide accurate estimation of certain features. While the estimation of covariance matrices is relatively stable, the expected return estimations are often showed to be imprecise. Biased parameter estimations translate to suboptimal allocation strategies. In particular, it is well-known that the mean-variance Markowitz portfolio optimization is sensitive to the return estimations, leading to unwise allocation decisions and poor performance when the return estimation is skewed.\\
\\
Stochastic optimization is another commonly adopted approach in portfolio optimization problems. Stochastic programming is a method for decision making under uncertainty with the underlying random process known or estimated. In a stochastic program, one takes action based on the information available at the time of decision, and is not allowed to take advantage of future information. In finance, it is called the non-anticipativity constraint. Examples of financial application of stochastic optimization include \cite{Mulvey2004}.\\
\\
In this paper, we tackle the portfolio allocation problem with an end-to-end approach where the model no longer depends on a prediction model that produces asset parameters to plug in the  portfolio model. The optimal asset allocation decisions are obtained from raw input data through a single data-decision pipeline embedded in a fully connected feed forward neural network. In this paper, we adopt both model-free and model-based learning approaches. 
Model-free portfolio learning method is based only on explicit layers. On the other hand, an implicit layer with a specified  optimization problem structure is integrated into the neural network in model-based methods.
\subsection{End-to-end Learning in Stochastic Optimization}
\cite{Donti2017} adopted the following framework in end-to-end stochastic optimization problems with a predictive model component. Let $(x \in \mathcal{X},y \in \mathcal{Y})\sim \mathcal{D}$ represents features and target variables that follows a distribution $\mathcal{D}$. The actions are denoted by $z \in \mathcal{Z}$ which incurs some expected loss $\mathcal{L}_{\mathcal{D}}(z) = \mathbb{E}_{x,y\sim \mathcal{D}}[c(x,y,z)]$ where $c$ is a nominal cost function. Optimal actions can be find by minimizing the loss function directly under the scenario where $\mathcal{D}$ is known. However, in practice true distribution is not known, and in the end-to-end framework conditional distribution $y|x$ is modeled by a parametric function $p(y|x;\theta)$. The goal is to find $\theta$ such that under $z^*(x;\theta)$ expected loss function is minimized
\begin{equation}
\begin{aligned}
    \underset{\theta}{\text{minimize}} \quad & \mathbb{E}_{x,y\sim \mathcal{D}}[c(x,y,z^*(x;\theta))]\\
    \text{s.t.} \quad & z^*(x;\theta) =\underset{z \in \mathbb{Z}}{\text{argmin }} \mathbb{E}_{y\sim p(y|x;\theta)} [c(x,y,z)]\\
\end{aligned}\label{end-to-end}
\end{equation}
where $\mathbb{Z}$ represents a constraint set that can have probabilistic and deterministic constraints. Notice that the cost function is the expected loss term which is the same function in the stochastic optimization problem. Therefore, the formulation is phrased as optimizing "task loss" in the end-to-end approach. Here the goal of the probabilistic model is not to find the best predictive outcome, but to find the best model that minimizes optimization problem's cost functions. \cite{Butler2021} incorporate this approach in the mean-variance portfolio optimization where asset returns are approximated with a linear model, and the cost function is the mean-variance term. One challenge in these integrated problems is the differentiating through the argmin operator. Specifically the gradient of the loss function needs to be computed $\frac{\partial\mathcal{L}}{\partial \theta} =\frac{\partial\mathcal{L}}{\partial z^*}\frac{\partial z^*}{\partial \theta} $ and the term $\frac{\partial z^*}{\partial \theta}$ requires argmin differentiation. To tackle this problem \cite{Donti2017} differentiate the Karush-Kuhn-Tucker (KKT) optimality conditions of the stochastic optimization problem (\ref{end-to-end}), and apply implicit function theorem which leads to set of equations that can be solved to obtain the necessary Jacobians. 
\subsubsection{End-to-end Portfolio Construction Approach}
In the portfolio problem, we utilize the end-to-end model approach in stochastic programming problems. Risk-budget portfolio method is incorporated as the underlying optimization model in the model-based portfolio learning.  The end-to-end portfolio model formulation differs from Problem (\ref{end-to-end}) in a few ways. In our portfolio optimization setting, the true labels are not available for the model parameters (risk budgets for each asset). Therefore, the loss function is not applicable here. Instead, similar to a reinforcement learning problem, we train the model based on a portfolio performance metric which we call the risk-reward function $f$. We evaluate the allocation decisions with some metric of the realized returns $r$. In particular, we optimize on a risk-reward function $f$ that encourages positive and stable returns, $\mathcal{R}_{\theta}(z) = \mathbb{E}_{x \sim \mathcal{D}}[f(x,r,z)]$. 

Let $\theta$ be the weights of the neural network that lead to a risk contribution decision, based on which allocation decision is made. Since there is not a closed-form expression for translating risk contribution to allocation\footnote{The closed form solution exists  when there are two assets ($n=2$). For the cases where $n>2$ numerical methods are used to find the optimal allocation.}, we need to employ an optimization layer to solve for the allocation. 
  We introduce the form of the optimization program in Section 3.3.1. For now, we denote the objective of this optimization layer by $c(x,z)$.
 We evaluate the allocation decisions with some metric of the realized returns. To mathematically present the formulation
\begin{equation}
\begin{aligned}
    \underset{\theta}{\text{minimize}} \quad & \mathbb{E}_{\mathcal{D}}[f(x,r,z^*(x;\theta))]\\
    \text{s.t.} \quad & z^*(x;\theta) =\underset{z \in \mathbb{Z}}{\text{argmin }} \mathbb{E}_{\mathcal{D}} [c(x,z)]\\
\end{aligned}\label{end-to-end-portfolio}
\end{equation}
\subsection{Neural Networks}
Artificial neural networks mimic the learning process of human brains to achieve certain assigned tasks. The usage of artificial neural networks traces back to last century (\cite{Sarle94neuralnetworks}), and recent years have witnessed growing attention of its applications in various fields including gaming, finance, biology, etc. In particular, with the assist of neural networks, researchers complete tasks that are intractable in the past. For example, \cite{casas} predicts the best performer out of three asset classes with neural network, and achieves satisfying return despite of lack of diversification. \cite{Mulvey2020} employ a deep neural network to learn a promising trading strategy for mean-reverting assets with linear transaction costs, when an advanced starting point is provided. They solve for the analytical solution under zero transaction costs, and find the optimal no-trade zone with the neural network. \cite{li2021} adopt a similar two-step procedure to tackle the portfolio allocation problem in a market with several possible regimes where linear transaction costs incur. A dynamic program offers the optimal allocation under zero transaction costs, which serves as the starting point for the neural network to look for the optimal trading strategy.\\
\\
An artificial neural network is an interconnected set of nodes, whose relationship is described by the weights. A typical feedforward neural network consists of an input layer, several hidden layers and an output layer.  The input layer collects the relevant information based on which the decisions are made. Each hidden layers contains a certain number of nodes, called neurons, whose value is a weighted sum of the neurons from the previous layer transformed by some activation function. Activation functions bring non-linearity into the system. Commonly used activation functions include ReLU (rectified linear unit), leaky ReLU, softmax, Sigmoid, Tanh, etc. The output layers provides the final result. Mathematically, let $x$ denote the inputs, $z_i$ denote the neurons in layer $i$ associated with weights $\theta_i$. Then, each layer is a linear transformation of the previous layer composed with an activation function. i.e., $z_{i+1} = \sigma(\theta_i z_i +b_i)$ where $\sigma$ is the activation function and $b_i$ is a bias term.\\
\\
Let $x$ be the set of input of the portfolio optimization problem, which may contain but is not restricted to historical returns and volatility. The set of weights, $\theta$, leads to an allocation decision $z$ based on the input, that translates to realized returns. We define the risk-reward function $\mathcal{R}$ to be a metric of the realized returns. Therefore, given the asset return dynamics $\mathcal{D}$, the risk-reward function $\mathcal{R}(z^*(x;\theta),r)$ is a function of inputs $x$, realized returns $r$ and weights $\theta$. The goal of the neural network is to find the optimal weights that optimize the risk-reward function $\mathcal{R}(z^*(x;\theta),r)$.\\
\\
The weights in the neural network are optimized via gradient descent. In each step, back-propagation process calculates the gradient of the risk-reward function with respect to the network parameters, $\frac{\partial\mathcal{R}}{\partial \theta}$, and then updated the weights with a pre-set learning rate. The learning terminates when the stopping criteria is met.

\subsection{Implicit Layers}
Deep learning models are built by stacking many layers together to create different architectures in order to solve a specific problem in hand. In feed-forward networks, a layer performs operations on the output of the previous layer, and provides output for the next layer. The majority of the layers in deep neural networks are defined explicitly, where  each neuron is a (possibly biased) linear combination of the previous layer composed with an activation function. Universal approximation theorem assures the ability of neural networks to approximate any well-behaved function, provided arbitrary width of the hidden layer. In real applications, on the other hand, it is not computationally practical to allow for arbitrary width. Further, when the data are not rich enough to represent the distribution, simply increasing the number of neurons can lead to severe overfitting. On the contrary, implicit layers are defined to satisfy some joint condition on the input and output and found to be useful in many areas: deep equilibrium models, differentiable optimization, neural ordinary differential equations as demonstrated in the tutorial by \cite{deepimplicitlayers}. In this paper, we employ an implicit layer in a neural network to embed, and differentiate the portfolio optimization problem.\\ 
\\
We introduce a type of layer whose values result from a convex optimization depending on the previous layer. In particular, the neuron values in the previous layer are used as parameters that characterize the convex optimization problem. Particularly a convex optimization layer can be defines as follows (\cite{Amos2019})
\begin{equation*}
    z_{i+1} = \underset{z}{\text{argmin }} f_{\theta}(z;z_i) \quad \text{s.t.} z \in \mathcal{S}_{\theta}(z_i)
\end{equation*}
where $z_i$ is the output of the previous layer and $z_{i+1}$ is the output of the convex optimization layer. The optimization model's objective function and constraint set are parametrized by $\theta$. The implicit function theorem is applied to convex $argmin$ operator to differentiate through the layer. Specifically, the gradients are obtained by applied the theorem with the KKT optimality conditions. More details on this procedure can be found in \cite{Amos2019} and \cite{deepimplicitlayers}. 
The optimal solution can be interpreted as a function of the parameters and is differentiable with respect to the neuron values in the previous layer, ensuring the back-propagation to work properly in a neural network.\\
\\
\texttt{CvxpyLayer} (\cite{Agrawal2019}) is a Python-embedded package that complies with the famous deep learning package \texttt{PyTorch} (\cite{Pytorch}), enabling one to build up a computational graph with handy back-propagation. The library is based on the convex optimization package \texttt{cvxpy} (\cite{diamond2016cvxpy}), and a convex optimization layer solves an optimization program whose solution depends on the previous layer. Such a layer may also add to the interpretability of neural network, where the relationship between the relevant consecutive layers are explained by the optimization program.\\
\\
When the convex optimization problem has a unique solution, the convex optimization layer is no different from a layer with deterministic functional relationship. The convex optimization layer, however, provides an elegant way to encode such relationship when analytical solution does not exist or is not tractable. In end-to-end model-based portfolio we implement risk budgeting optimization problem as a convex optimization layer in the network.

\subsubsection{Risk-budgeting Portfolio Model}
We consider the following portfolio models in this paper: (i) a model-free approach where the neural network learns the allocation strategy from raw input directly, and (ii) a model-based strategy where the neural network learns the risk budget from the input, accordingly to which the allocation is decided. The traditional portfolio choice model is introduced by \cite{Markowitz1952} that results the optimal allocation decision based on a mean-variance criteria. Even though mean-variance portfolio theory is widely accepted among academics and industry practitioners, it faces practical drawbacks, especially in the prediction of the asset parameters.  \cite{Chopra1993} emphasize the  importance of the asset return estimations for investors that are using mean-variance framework to allocate their wealth. 
Obtaining good return forecasts is a hard problem, and even small errors can lead to substantial difference in the outcome. Risk-budgeting portfolio optimization is introduced mathematically by \cite{Bruder2012}, and the special case equal risk contribution portfolio is appeared formally in \cite{Maillard2010}. It produces allocations based the idea of risk-based diversification. Unlike mean-variance portfolio model, the risk budgeting portfolio problem does not require asset return forecasts as input, and it is robust misspecifications of in covariance matrix (\cite{Ardia2017}).  In this subsection, we will introduce the risk budgeting allocation model.\\
\\
First we define the risk contribution of each asset in a portfolio. Suppose there are $n$ assets being considered in a portfolio, and the covariance matrix is $\Sigma$. Given an allocation $x$, the risk contribution of an asset is a measure of how much it contributes to the resulting portfolio. In particular, when using volatility as the risk measure, the risk contribution of asset $i$ is $RC_i = \frac{x_i(\Sigma x)_i}{\sqrt{x^T\Sigma x}}$. Note that the risk contributions from all assets sum up to one.\\
\\
A risk budgeting portfolio is one where the allocation is made so that the contribution of risk from each asset matches the pre-defined risk budget. A special case of risk budgeting is risk-parity, where each asset has the same degree of risk contribution. Risk parity portfolios have gain increasing attention in recent years. Compared to the traditional mean-variance Markowitz approach, the risk parity portfolio does not depend on return estimations, and provide a more robust performance over different market circumstances, often leading to a higher Sharpe ratio. On the other hand, some critics point out that risk-parity could be sensitive to the underlying asset universe. Since risk-parity focuses solely on volatility, the portfolio can be hurt by assets with low or negative returns, let alone if an asset has negative return and low volatility. A general risk budgeting portfolio can potentially mitigate the drawback by allocating less risk budget to the undesired asset. It also allows for more flexibility in accommodating investors' risk preferences.\\
\\
Suppose $b\in\mathbb{R}^n$ with $\Sigma_i b_i = 1, b\geq 0$ is the risk budget vector, whose entries are risk contribution allocated to each asset. A long-only risk budgeting portfolio can be found with an optimization problem suggested by \cite{Bruder2012}:
\begin{align}
\begin{split}
    \underset{x}{\text{minimize}} \hspace{1cm} & \left( \frac{x_i(\Sigma x)_i}{\Sigma_j x_j(\Sigma x)_j} - b_i \right) ^2\\
    s.t. \hspace{1cm}  & \sum_{i=1}^n x_i = 1\\
    & x\geq0.
\end{split}
\label{rb_nonconvex}
\end{align}
Problem (\ref{rb_nonconvex}) is not a convex program. Fortunately, there exist equivalent ways to formulate risk budgeting portfolio optimization as a convex program:
\begin{align}
\begin{split}
    \underset{y}{\text{minimize}} \hspace{1cm} & \sqrt{y^T\Sigma y}\\
    s.t. \hspace{1cm}  & \sum_{i=1}^n b_i ln(y_i) \geq c\\
    & y\geq0.
\end{split}
\label{rb_log}
\end{align}
where $c$ is an arbitrary positive constant. The resulting portfolio can be found with normalizing $x_i = \frac{y_i}{\Sigma y_i}$. Equivalence of Problem \ref{rb_nonconvex} and \ref{rb_log} can be verified with KKT conditions (\cite{Maillard2010}). We consider only long positions in this paper due to its nice mathematical properties. \cite{Bai2016} show that there are multiple solutions when shorting is allowed in risk parity portfolios. For a portfolio with $n$ assets, there are $2^{n-1}$ risk parity allocations. One need to enumerate all these solutions and find the optimal allocation according to their preference. \cite{Richard2019} discuss constrained risk budgeting portfolios and show that with addition of weight constraints,  ex-ante and ex-post risk budgets are not same. They claim that when weight constraints added to the optimization Problem \ref{rb_log}, risk budget choices have little impact, and portfolio allocation is mainly driven by the constraints. Due to these problems we do not incorporate leverage or specific weight constraints, but this area can be explored in the  future research.   
\subsubsection{Differentiating through Risk-budget Layer}
With chain rule, the gradient of risk-reward function in terms of neural network weights $\theta$ can be expressed as $\nabla_\theta\mathcal{R}=\frac{\partial\mathcal{R}}{\partial z^*} \frac{\partial z^*}{\partial y^*} \frac{\partial y^*}{\partial b} \frac{\partial b}{\partial\theta}$, where $b$ is the risk budget provided by the neural network, and $y^*$ is the optimal solution from Problem (\ref{rb_log}) with $z^* = \frac{y^*}{\|y^*\|_1}$. The derivative $\frac{\partial z^*}{\partial y^*}$ is obtained from the relationship, and $\frac{\partial b}{\partial\theta}$ can be found with standard neural network back-propagation. Optimizing the neural network is now left with differentiating through the convex optimization layer, which in our case, the risk-budget layer. In this subsection, we describe how one finds $\frac{\partial y^*}{\partial b}$.\\
\\
We will provide the KKT condition of the optimization problem and derive the derivative $\frac{\partial y^*}{\partial\theta}$ implicitly. Let $\lambda\in\mathbb{R}$ and $\mu\in\mathbb{R}^n$ be the dual variables corresponding to the two constraints in Problem (\ref{rb_log}), respectively. First we will derive the Langrangian of the Problem (\ref{rb_log}) 
\begin{equation}
\begin{aligned}
\mathcal{L}(y,\lambda,\mu) = y^T\Sigma y +\lambda (c- \sum_{i=1}^n b_i y_i)+\mu^Ty\\
\end{aligned}
\end{equation}
\noindent
Following primal and dual feasibility, stationarity, and complementary slackness conditions, we write out the KKT conditions as below:
\begin{align}
\begin{split}
    2\Sigma y - \lambda b^T\frac{1}{y}+\mu = 0\\
    b^Tln(y)\geq c\\
    y \geq 0\\
    \lambda \geq 0 \\
    \mu\geq 0\\
    \lambda(c-b^T ln(y))=0\\
    \mu^T y =0
\end{split}
\end{align}
where $\frac{1}{y}$ is the vector where each entry is the inverse of corresponding entry in vector $y$, and $ln(y)$ is the vector where each entry is logarithm of corresponding entry in vector $y$.\\
\\
Taking the differentials of the equalities in the KKT condition, we get
\begin{equation}
\begin{bmatrix}
2\Sigma+(\lambda b^T\frac{1}{y^2})I & -(b^T\frac{1}{y})I & I\\
-\lambda(b^T\frac{1}{y}) & c-b^T ln(y) & 0\\
\mu & 0 & y
\end{bmatrix}
\begin{bmatrix}
dy\\
d\lambda\\
d\mu
\end{bmatrix}
=
\begin{bmatrix}
-2d\Sigma y + \lambda db^T\frac{1}{y}\\
-\lambda(dc - db^Tln(y))\\
0
\end{bmatrix}
\end{equation}
where $I$ is the identity matrix. The system of equations implies the derivative of optimal $y^*$ with respect to weights of the risk contributions $b$. To calculate $\frac{\partial y^*}{\partial b}$, we set $db = I$ and all other differential terms as zero, and solve the linear system for $dy$. After each training step, the weights $\theta$ are updated with $(\theta - \text{learning rate}*\nabla_\theta\mathcal{R})$, until the stopping criteria is met. 
\subsection{Model Architecture}
\subsubsection{Model-free}
\begin{figure}[hbt!]
    \centering
    \includegraphics[width=\textwidth,center]{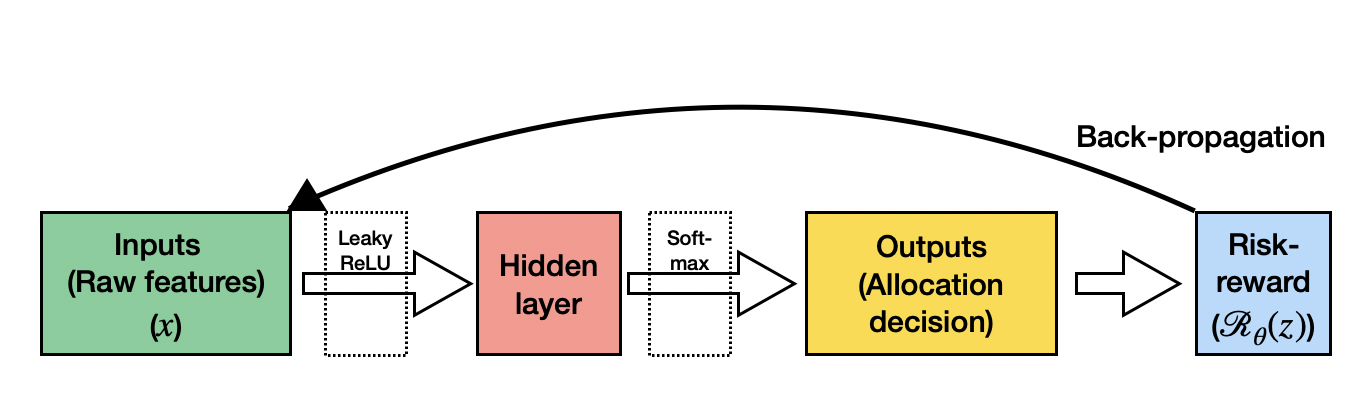}
    \caption{Computational graph of model-free approach.}
    \label{fig:computational_graph_free}
\end{figure}

For the model-free approach, we employ a fully feed-forward neural network with one hidden layer. The input layer consists of raw features, and the output layer consists of $n$ neurons, representing the allocation in each asset. The computational graph appears in Figure \ref{fig:computational_graph_free}.\\
\\
\subsubsection{Model-based}
\begin{figure}[hbt!]
    \centering
    \includegraphics[width=\textwidth,center]{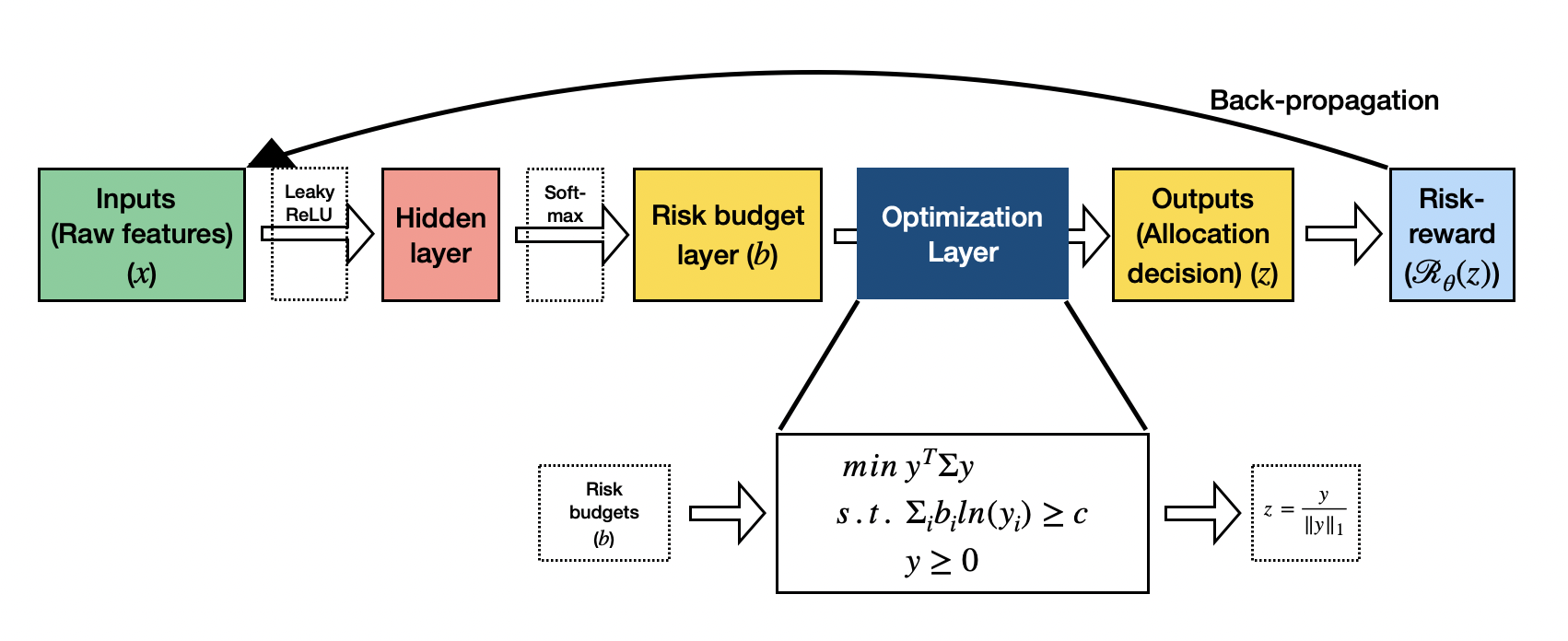}
    \caption{Computational graph of model-based approach.}
    \label{fig:computational_graph_based}
\end{figure}

For the risk-budget-based strategy, the computational graph of our neural network appears in Figure \ref{fig:computational_graph_based}. The neural network consists of four layers, in the order of computation: an input layer, two hidden layers, and a convex optimization layer, where the last layer provides the asset allocation decision as the output.\\
\\
The input layer takes in raw features, including historical returns and volatility of each asset. To compile with non-anticipativity constraint, the input feature does not include asset returns on the day of interest. The hidden layers are ordinary fully connected layers whose neuron values come from a linear transformation of the input layer, and then composed with an activation function. The second hidden layer has the number of neurons equalling the number of assets. We apply a softmax function on the second hidden layer to normalize the values and interpret them as the risk budget allocated to each of the assets. In the convex optimization layer, we solve a convex optimization that translates the risk budgets into allocations. Assuming uniqueness and accuracy of the optimal solution, there are no weights to be learned nor randomness involved in the convex optimization layer. \\
\\
When training the neural network, with the realized asset returns on the day of interest, we are able to calculate the actual return of the portfolio given the allocation suggested by the neural network. The loss function is defined with the returns. We consider a various set of loss functions and test their out-of-sample performance, including Sharpe ratio, mean-variance objective, average drawdown, etc.\\
\\
To address the fact that financial data are often non-stationary, we train the neural network on a rolling-basis. In particular, in order to learn the weights that apply to day $t$'s allocation, we train the neural network based on data of past $K$ days, namely, from day $t-K$ to day $t-1$. The algorithm is described in Algorithm \ref{algo}.\\

\begin{algorithm}[H]\label{algo}
\SetKwInOut{Input}{Input}
\SetKwInOut{Output}{Output}
\Input{past 5 daily returns and past 10,20,30 day average returns and volatilities of each asset}
\Output{Portfolio performance over a period from day $K$ to day $T$}
\For{each day $t\in\{K,K+1,...,T\}$}{
Initialization: weights $\theta$;\\
\For{each training step}{
    \For{each day $s\in\{t-K,t-K+1,...,t-1\}$}{
        Use $\theta$ and features $x_s$ to calculate the risk contribution $b_s$ on day $s$\\
        Calculate the estimated covariance matrix $\Sigma_s$ for day $s$\\
        $y_s = \text{argmin } y^T\Sigma_s y \hspace{0.5cm} \text{ s.t. } b_s^T ln(y)\geq c \text{ and } y\geq 0$\\
        $z_s = \frac{y_s}{\|y_s\|_1}$ is the allocation on day $s$\\
        Record the realized return $r_s$ on day $s$ with actual returns
    }
    Calculate risk-reward function $\mathcal{R}_\theta(z)$\\
    Back-propagate: $\theta \leftarrow (\theta - \text{learning rate}*\nabla_\theta\mathcal{R})$
}
   With trained $\theta^*$, calculate risk contribution on day $t$ and allocate accordingly\\
   Record the realized return on day $t$
}
\caption{Calculation of portfolio performance}
\end{algorithm}

\subsection{Computational Set-up}
\subsubsection{Software}
Experiments are written in \texttt{Python} programming language and two main software libraries are To construct neural networks we utilize from \texttt{PyTorch (1.8.1)}, and \texttt{CvxpyLayers (0.1.4)} is used to construct an implicit layer to represent  portfolio optimization problem as an individual layer in the network. The experiments are run on the Princeton Tiger Research cluster with Intel Skylake CPU nodes.
\subsubsection{Training Neural Network}
In the model-free approach, we adopt one hidden layer with leaky ReLU activation function. The number of neurons in the output layer is the same as number of assets in the portfolio which represents the allocation decision. Softmax activation function is applied to ensure the allocation adds up to one.\\
In the model-based approach, we adopt one hidden layer with leaky ReLU activation function. The next layer represents the risk contributions (or risk budgets), where we employ softmax activation function to ensure the risk contributions add up to one. Then, an implicit layer solves for the allocation decision.\\
In both approaches, the leaky ReLU parameter is chosen to be $\alpha=0.1$. i.e, $\sigma_{ReLU}(x) = max(x,0) + 0.1min(x,0)$. Two risk-reward functions are chosen to train neural networks: sharpe ratio and cumulative return of the portfolio over the training period.  
\section{Simulation Study}\label{sec:sim-study}
\subsection{Set-up}
We simulate a seven-asset universe where the returns of the assets follow a multi-variate normal distribution, and is independently and  identically distributed for each trading day. To mimic real market environment, the distribution parameters are determined by the mean and covariance matrix of daily returns of seven ETFs from 2011 to 2021: VTI, IWM, AGG, LQD, MUB, DBC and GLD, the details of which are described in Section \ref{sec:market-data}. The expected daily returns for the seven assets in the simulation are 0.059\%,  0.013\%, -0.011\%,  0.022\%,  0.056\%, 0.017\% ,  0.017\%, respectively.

\subsection{Results}
We run the following strategies on simulated data of length 175 days: model-based end-to-end learning,
model-free end-to-end learning, and the nominal risk parity strategy as a benchmark. For the end-to-end neural networks, the hyperparameters are chosen as below:
\begin{itemize}
    \item Number of neurons in the hidden layer: 32
    \item Learning rate: 10
    \item Number of steps: 50
    \item Rolling window for training purpose: 150 days
    \item Test window: 5 days
\end{itemize}
For every 5-day period, we train the neural network with the data of 150 days immediately previous to the period of interest. We keep the same weights for the 5-day period, and repeat the same process for the next period. To test the robustness of performance with each end-to-end method, we run 100 seeds on both, and plot the best performing seed, worst performing seed, as well as the median and average results.

\begin{figure}[hbt!]
    \centering
     \begin{subfigure}{0.48\textwidth}
         \centering
         \includegraphics[width=1.15\textwidth,center]{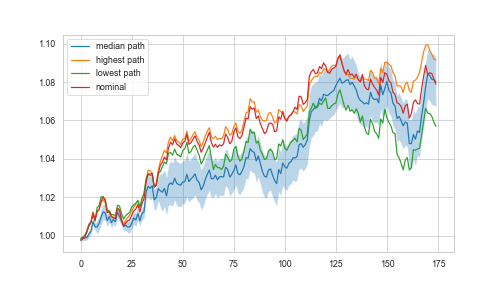}
         \caption{Model-based}
     \end{subfigure}
     \begin{subfigure}{0.48\textwidth}
         \centering
         \includegraphics[width=1.15\textwidth,center]{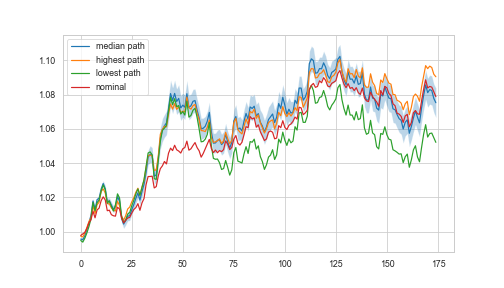}
         \caption{Model-free}
     \end{subfigure}
    \caption{Computational result on simulated data when the tuning objective is chosen to be Sharpe ratio.}
    \label{fig:sim_sr}
\end{figure}
\noindent
To analyze the performance of model-based and model-free methods compared to the risk-parity benchmark, we propose the following hypotheses and see if we have strong enough evidence to reject them. To quantify the performance, we adopt geometric average return over average drawdown as the main metric. We train the neural network with Sharpe ratio as the tuning objective function.

\begin{hyp}\label{hyp_free_vs_based}
The average performance (in terms of geometric average return over average drawdown) of the model-free end-to-end method is no less than that of the model-based method.
$$H_0: \bar{R}_{free}\geq\bar{R}_{based} \hspace{1cm}\text{ vs. }\hspace{1cm} H_a: \bar{R}_{free}<\bar{R}_{based} $$
\end{hyp}
Based on the 100 random seeds, the average performance of model-free method in terms of geometric average return over average drawdown is 7.877 with standard deviation 1.522. That of the model-based method averages at 18.997 with standard deviation 3.941. Applying the one-sided hypothesis testing, the test statistic $Z = \frac{7.877 - 18.977}{\sqrt{\frac{1.522^2}{100}} + \frac{3.941^2}{100}} = -26.3$. We have sufficient evidence to conclude that the model-based method outperforms the model-free method, at 1\% significance level.\\
\\
Similar conclusion is reached if we adopt Sharpe ratio as the performance metric.

\begin{figure}[hbt!]
    \centering
     \begin{subfigure}{0.48\textwidth}
         \centering
         \includegraphics[width=1.15\textwidth,center]{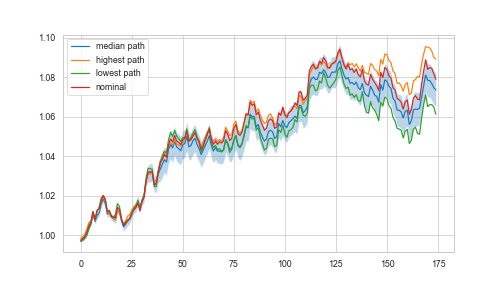}
         \caption{Model-based}
     \end{subfigure}
     \begin{subfigure}{0.48\textwidth}
         \centering
         \includegraphics[width=1.15\textwidth,center]{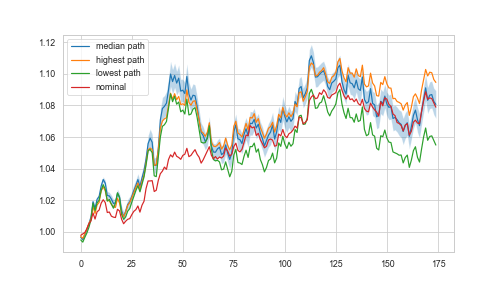}
         \caption{Model-free}
     \end{subfigure}
    \caption{Computational result on simulated data when the tuning objective is chosen to be cumulative return.}
    \label{fig:sim_ret}
\end{figure}

\begin{hyp}\label{hyp_benchmark_vs_based}
The average performance (in terms of geometric average return over average drawdown) of the model-based method is no greater than that of risk-parity strategy.
$$H_0: \bar{R}_{based}\leq{R}_{parity} \hspace{1cm}\text{ vs. }\hspace{1cm} H_a: \bar{R}_{based}>{R}_{parity} $$
\end{hyp}
Applying nominal risk parity on the simulated dataset leads to a geometric average return over average drawdown of 16.901. The test statistic is $Z = \frac{18.977-16.901}{\frac{3.941}{\sqrt{100}}} = 5.27$. We have sufficient evidence to conclude that the model-based method on average outperforms the nominal risk parity, at 1\% significance level.\\
\\
Similar conclusion is reached if we adopt Sharpe ratio as the performance metric, where the test statistic $Z=\frac{2.278-2.182}{\frac{0.289}{\sqrt{100}}}=3.32$. In particular, 60 out of 100 seeds provides higher Sharpe ratio on the dataset than nominal risk parity.\\
\\
In addition, we provide the statistics when cumulative return is used as the tuning objective for neural network training. For Hypothesis (\ref{hyp_free_vs_based}), the test statistic is $Z = \frac{6.321 - 17.355}{\sqrt{\frac{1.212^2}{100}} + \frac{3.084^2}{100}}=-33.30$, and we have sufficient evidence to reject the null hypothesis and conclude the superiority of the model-based method at 1\% significance level. For Hypothesis (\ref{hyp_benchmark_vs_based}), the test statistics is $Z = \frac{17.355-16.901}{\frac{3.084}{\sqrt{100}}} = 1.47$. There is evidence that the model-based approach beats nominal risk parity, at 10\% significance level. Figure \ref{fig:sim_ret} shows the performance of various seeds when trained on cumulative return.
\clearpage
\section{Real Market Data}\label{sec:market-data}
We use daily returns of seven exchange-traded funds (ETFs) instead of individual assets to represent stock, bond and commodity market conditions: VTI (Vanguard Total Stock Market ETF), 
IWM (iShares Russell 2000 ETF), AGG (iShares Core U.S. Aggregate Bond ETF), LQD ( iShares iBoxx Investment Grade Corporate Bond ETF), 
MUB (iShares National Muni Bond ETF), DBC (Invesco DB Commodity Index Tracking Fund), and GLD (SPDR Gold Shares). ETF performance statistics over the time period 2011-2021 are presented in Table \ref{tab:asset}.  
\begin{table}[hbt!]
    \centering
\begin{tabular*}{\textwidth}{c@{\extracolsep{\fill}} cccccc}
\toprule
{\textbf{ETFs}} &   Return & Volatility &   Sharpe &     MDD & Calmar Ratio & Return/Ave.DD \\         \\
\midrule
VTI    &   0.1410 &     0.1759 &   0.7677 &  0.3500 &       0.3849 &        4.1473 \\
IWM    &   0.1248 &     0.2194 &   0.5418 &  0.4113 &       0.2767 &        1.7915 \\
AGG    &   0.0335 &     0.0401 &   0.6995 &  0.0958 &       0.2913 &        2.1182 \\
LQD    &   0.0537 &     0.0724 &   0.6650 &  0.2176 &       0.2209 &        2.1832 \\
MUB    &   0.0417 &     0.0514 &   0.7052 &  0.1368 &       0.2642 &        1.8995 \\
DBC    &  -0.0477 &     0.1611 &  -0.3273 &  0.6614 &      -0.0777 &       -0.1374 \\
GLD    &   0.0204 &     0.1584 &   0.0951 &  0.4556 &       0.0330 &        0.0561 \\
\bottomrule
\end{tabular*}
    \caption{Annualized ETF performance statistics over the period 2011-2021.}
    \label{tab:asset}
\end{table}

\subsection{Neural Network Training and Hyperparameter Selection}\label{subsec:hyperparameter}
We present the computational results over the last ten years of daily data, where we keep the first six years (2011-2016/12) for hyperparameter search and the remaining years (2017-2021) for out-of-sample testing. We construct two neural networks: model-based and model-free end-to-end portfolios with a fully connected neural network consisting of a single hidden layer. In the backtesting framework, the neural network models are re-trained every 25 days with a look-back window of 150 days. 
To keep the model-based network less complicated, we use a sample estimator of the covariance matrix in the end-to-end risk budgeting and nominal risk parity portfolios obtained from the historical returns of the past 30 days.  Exploring other estimation techniques or incorporating covariance matrix as a network parameter can be interesting future research directions. 
We split the in-sample period into training (2011-2014/12), and validation (2015-2016/12) sets to choose hyperparameters of the network. As the universal approximation theorem establish the ability of shallow feed-forward neural networks to approximate any continuous function, we choose to avoid unnecessary complexity and to go with a relatively shallow network consisting of 32 neurons in a single layer. We carefully tune the learning rate ($\eta$) and the number of training steps ($n$). 
A higher learning rate provides faster learning in a few steps. However, it might lead to zigzagging due to large steps. Whereas low learning rates have smoother convergence paths, but learning is much slower and with the downside of getting stuck at a local optimum point. We use linear update rules for learning rates and decrease it by the factor of 0.9 in every three steps.  The results over the range $\eta \in \{50,100,150,250,300,500\}$ and $n \in \{ 5,10,25,50\}$ are presented in Table \ref{tab:hyperparameter-Sharpe} and \ref{tab:hyperparameter-Return}. Ideally, an optimal hyperparameter choice should provide reasonably good performance in both training and validation sets.  Notice that as training steps increase, training time per batch increases significantly\footnote{We believe that long running time is due to available solvers in the \texttt{CvxpyLayer} library at that time.}. In order to select the hyperparameters, we use the following procedure: First, filter the parameters sets that are in the top half of both train and validation sets. Then choose the one with the best performance in the validation set to guarantee the generalization. We find that for Sharpe-based portfolios, the learning rate of 150 provides good performance with ten training steps. Optimal parameter set for cumulative return function found to be learning rate of 300 with 25 training steps. 

\begin{table}[]
\centering
\begin{tabular*}{\textwidth}{p{1.5cm}@{\extracolsep{\fill}} p{1cm}p{2.5cm}p{2.5cm}p{2cm}}
\toprule
 learning rate ($\eta$)&  number of steps &  train set performance &  validation set performance &   time per training batch (min.) \\
\midrule
            50 &                5 &             0.9787 &                  0.4153 &   1.59 \\
           100 &                5 &             1.0331 &                  0.3739 &   1.58 \\
           150 &                5 &             0.9760 &                  0.4399 &   1.42 \\
           200 &                5 &             0.7501 &                 -0.1497 &   1.61 \\
           300 &                5 &             0.9888 &                  0.6413 &   1.55 \\
           500 &                5 &             0.8164 &                 -0.0293 &   1.44 \\
\midrule
            50 &               10 &             1.0386 &                  0.3697 &   2.78 \\
           100 &               10 &             0.8861 &                  0.1702 &   2.95 \\
           \textbf{150} &               \textbf{10} &             \textbf{1.1842} &                  \textbf{0.9892} &   \textbf{2.83} \\
           200 &               10 &             1.0707 &                 -0.6818 &   3.10 \\
           300 &               10 &             1.3672 &                  0.8938 &   3.21 \\
           500 &               10 &             0.5427 &                  0.6792 &   3.00 \\
\midrule
            50 &               25 &             1.1667 &                  0.5162 &   7.04 \\
           100 &               25 &             0.8921 &                  0.7556 &   7.37 \\
           150 &               25 &             1.1419 &                  0.4853 &   7.16 \\
           200 &               25 &             1.3760 &                 -0.4079 &   7.22 \\
           300 &               25 &             1.1923 &                  0.3381 &   8.84 \\
           500 &               25 &             0.2712 &                  0.1384 &   8.43 \\
\midrule
            50 &               50 &             1.2855 &                  0.7015 &  14.80 \\
           100 &               50 &             1.0202 &                  0.5708 &  14.84 \\
           150 &               50 &             1.2534 &                  0.5965 &  15.12 \\
           200 &               50 &             1.0167 &                 -0.4878 &  16.28 \\
           300 &               50 &             0.7900 &                  0.4576 &  18.27 \\
           500 &               50 &             0.2073 &                  0.0693 &  17.29 \\
\bottomrule
    \end{tabular*}
    \caption{Train and validation set performance metrics (in terms of Sharpe ratio) for end-to-end portfolios that are trained with Sharpe ratio.}
    \label{tab:hyperparameter-Sharpe}
\end{table}

\begin{table}[]
\centering
\begin{tabular*}{\textwidth}{p{1.5cm}@{\extracolsep{\fill}} p{1cm}p{2.5cm}p{2.5cm}p{2cm}}
\toprule
 learning rate ($\eta$)&  number of steps &  train set performance &  validation set performance &   time per training batch (min.) \\
\midrule
       50 &                5 &             1.1289 &                  1.0288 &   1.59 \\
           100 &                5 &             1.1237 &                  1.0243 &   1.58 \\
           150 &                5 &             1.1497 &                  1.0211 &   1.42 \\
           200 &                5 &             1.1458 &                  1.0279 &   1.61 \\
           300 &                5 &             1.0662 &                  1.0236 &   1.55 \\
           500 &                5 &             1.0897 &                  1.0110 &   1.44 \\
\midrule
            50 &               10 &             1.1407 &                  1.0240 &   2.78 \\
           100 &               10 &             1.1174 &                  1.0231 &   2.95 \\
           150 &               10 &             1.1612 &                  1.0095 &   2.83 \\
           200 &               10 &             1.0918 &                  1.0135 &   3.10 \\
           300 &               10 &             1.0874 &                  1.0246 &   3.21 \\
           500 &               10 &             1.1060 &                  1.0306 &   3.00 \\
\midrule
            50 &               25 &             1.1412 &                  1.0229 &   7.04 \\
           100 &               25 &             1.1247 &                  1.0246 &   7.37 \\
           150 &               25 &             1.1626 &                  1.0183 &   7.16 \\
           200 &               25 &             1.1793 &                  1.0364 &   7.22 \\
           \textbf{300} &               \textbf{25} &             \textbf{1.1427} &                  \textbf{1.0671} &   \textbf{8.84 }\\
           500 &               25 &             1.1143 &                  0.9910 &   8.43 \\
\midrule            
            50 &               50 &             1.1357 &                  1.0233 &  14.80 \\
           100 &               50 &             1.1354 &                  1.0240 &  14.84 \\
           150 &               50 &             1.1652 &                  1.0214 &  15.12 \\
           200 &               50 &             1.1975 &                  1.0540 &  16.28 \\
           300 &               50 &             1.1178 &                  1.0865 &  18.27 \\
           500 &               50 &             1.1297 &                  0.9748 &  17.29 \\
\bottomrule
    \end{tabular*}
    \caption{Train and validation set performance metrics (in terms of cumulative return) for end-to-end portfolios that are trained with cumulative return.}
    \label{tab:hyperparameter-Return}
\end{table}
\subsection{End-to-end Risk Budgeting Portfolios}
\begin{figure}[hbt!]
    \centering
    \begin{subfigure}{0.9\textwidth}
    \centering
    \includegraphics[width=1.15\textwidth,center]{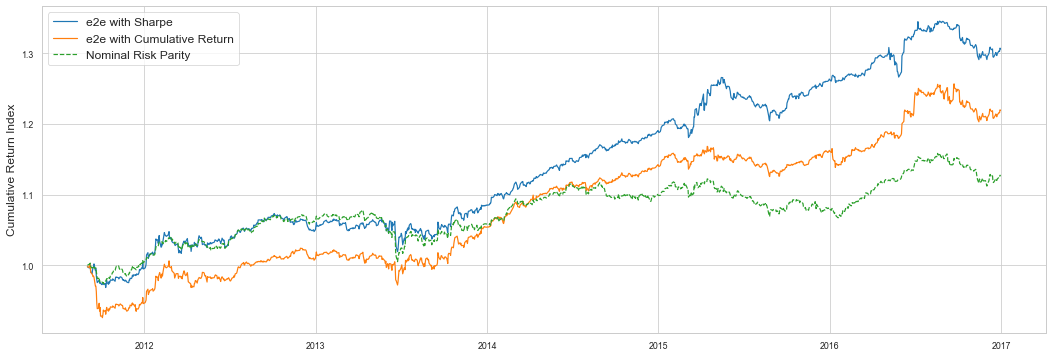}
    \end{subfigure}
    \begin{subfigure}{0.9\textwidth}
    \centering
        \includegraphics[width=1.15\textwidth,center]{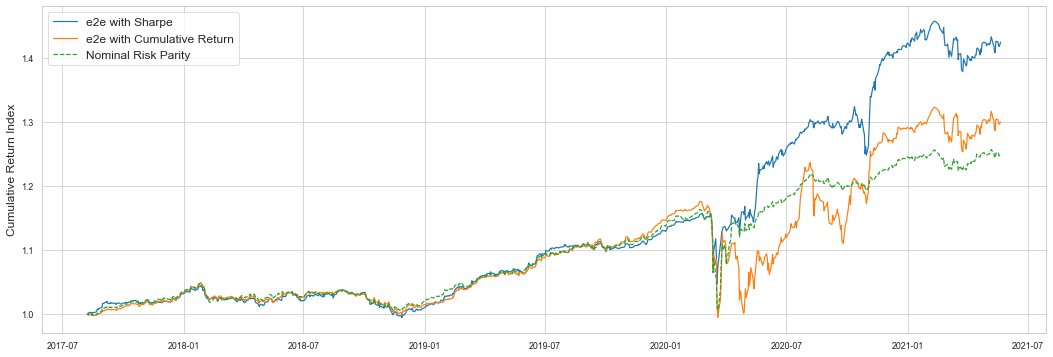}
    \end{subfigure}
    \caption{Optimal end-to-end (e2e) portfolio performances over in-sample (top) and out-of-sample (bottom) periods.}
    \label{fig:insample-and-outofsample-e2e}
\end{figure}

\begin{figure}[hbt!]
    \centering
    \begin{subfigure}{0.9\textwidth}
    \centering
        \includegraphics[width=1.15\textwidth,center]{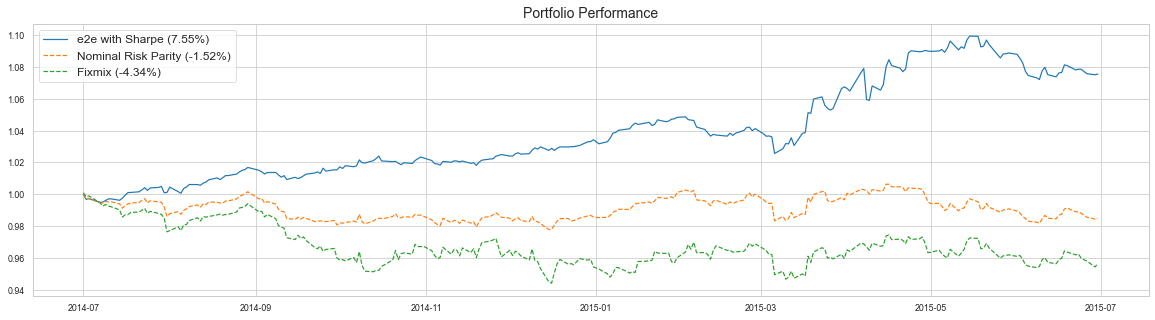}
    \end{subfigure}
    \begin{subfigure}{0.9\textwidth}
    \centering
    \includegraphics[width=1.15\textwidth,center]{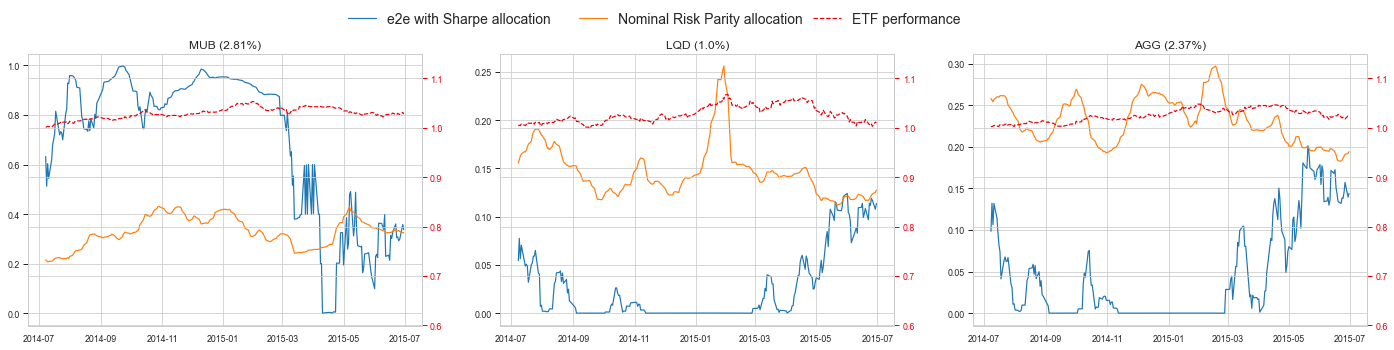}
    \end{subfigure}
        \begin{subfigure}{0.9\textwidth}
    \centering
    \includegraphics[width=1.15\textwidth,center]{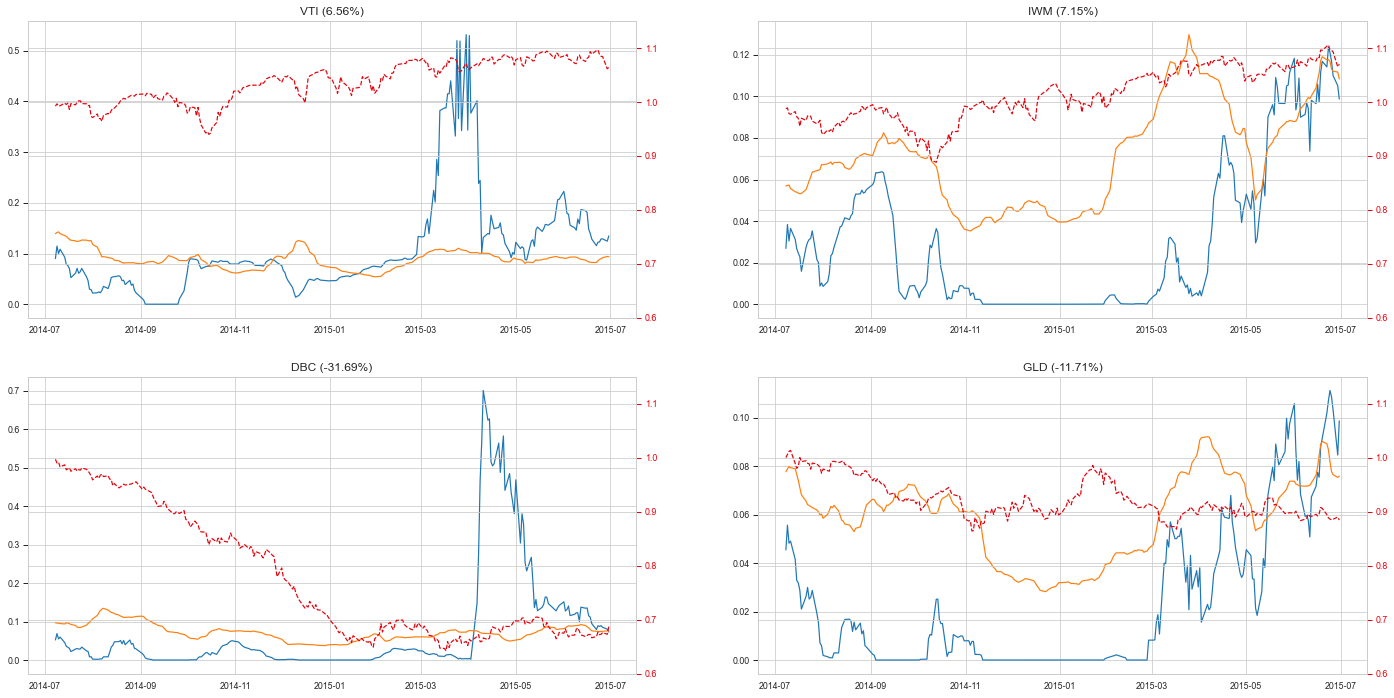}
    \end{subfigure}
    \caption{Portfolio performances and allocations over the in-sample period 2014/07-2015/07. Reported asset allocations are weekly rolling averages. Numbers in the parenthesis show annualized returns.}
    \label{fig:after2013}
\end{figure}

\begin{figure}[hbt!]
    \centering
        \begin{subfigure}{0.9\textwidth}
    \centering
    \includegraphics[width=1.15\textwidth,center]{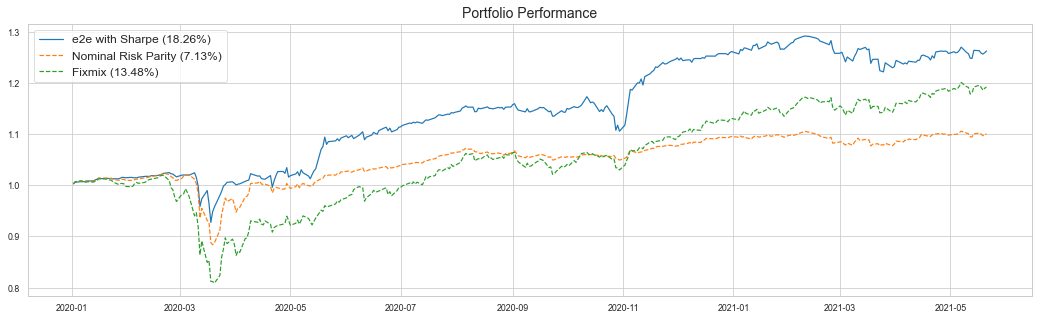}
    \end{subfigure}
        \begin{subfigure}{0.9\textwidth}
    \centering
    \includegraphics[width=1.15\textwidth,center]{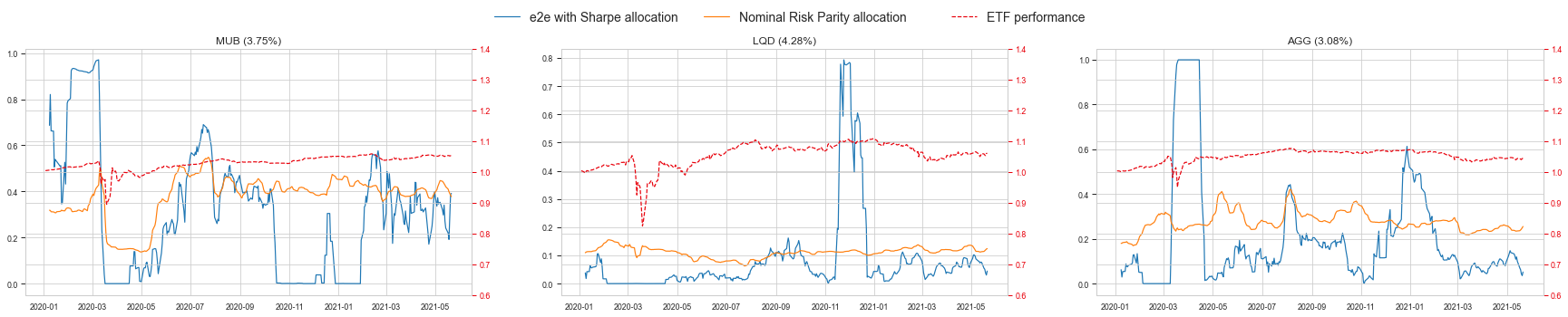}
    \end{subfigure}
        \begin{subfigure}{0.9\textwidth}
    \centering
    \includegraphics[width=1.15\textwidth,center]{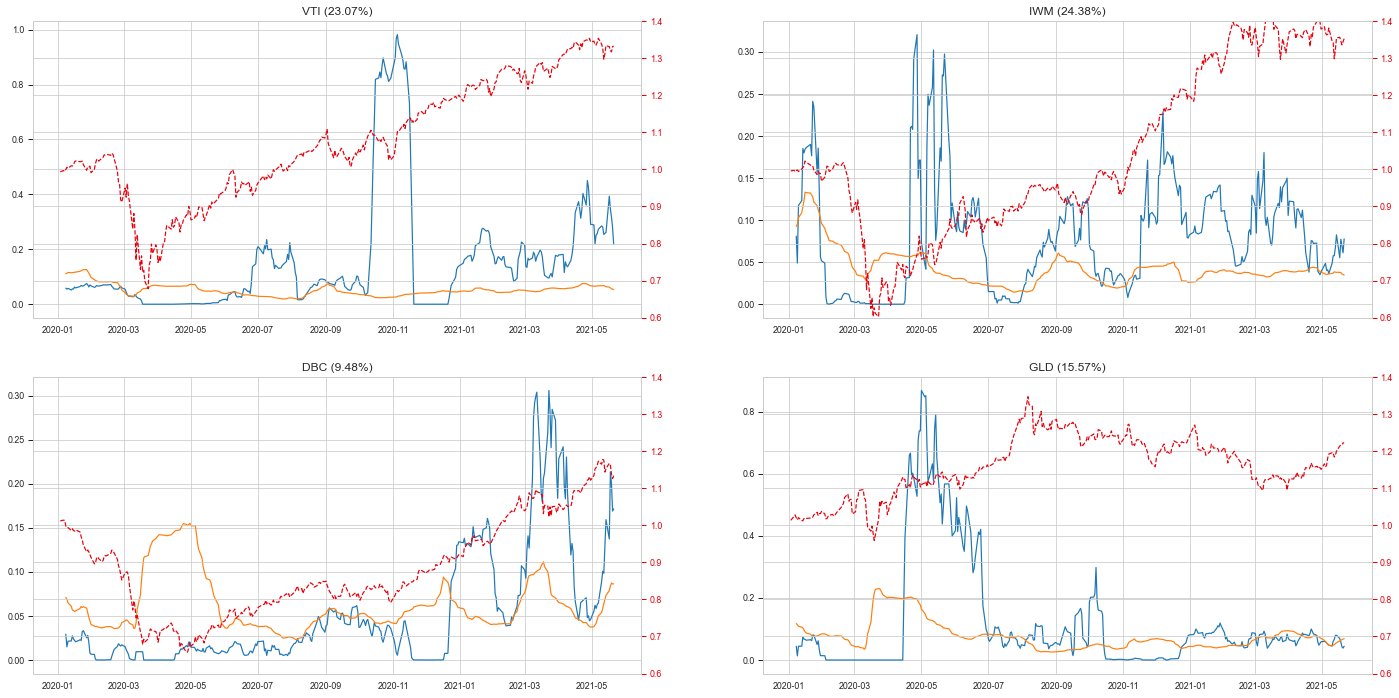}
    \end{subfigure}
    \caption{Out-of-sample portfolio performances after the downturn in March 2020. Reported asset allocations are weekly rolling averages. Numbers in the parenthesis show annualized returns.}
    \label{fig:aftercovid}
\end{figure}

We compare end-to-end approach with a nominal risk parity portfolio. End-to-end portfolios have dynamic risk budget allocations for each asset based on learning results, whereas risk parity portfolio distributes risk contribution evenly across all underlying assets.
With the end-to-end model risk budgets can be adjusted based on assets' performance which helps to enhance the portfolio performance over nominal allocation. Figure \ref{fig:insample-and-outofsample-e2e} demonstrates cumulative performances of end-to-end and nominal portfolios, and Table \ref{tab:e2e-stats} presents annualized portfolio statistics over the in-sample and out-of-sample periods. Fix-mix with equal weights (1/N) is showed by many researchers to be a robust and relatively strong benchmark. In this data set, the fix-mix strategy has worse risk-adjusted performance than risk-based asset allocation techniques. In both time periods, Sharpe-based end-to-end portfolio outperforms nominal risk-parity portfolio in terms of risk-adjusted returns, and the difference becomes more obvious after the crash of the nominal portfolio in 2013 and 2020. 
During in-sample period (2011-2016/12), Sharpe based end-to-end model achieves a Sharpe ratio of 1.11 whereas nominal risk parity has 0.62. End-to-end model trained with cumulative return function achieves higher cumulative wealth than nominal portfolio ($1.22>1.13$), and has better risk-adjusted performance in terms of Sharpe ratio ($0.80>0.63$). Fix-mix strategy has the worst risk-adjusted performance with a Sharpe ratio of 0.41. The performance difference between end-to-end Sharpe based model and the nominal risk parity over the in-sample period is evident in terms of the return to average drawdown ratio ($4.56>1.54$). However, over the out-of-sample period (2017-2021), performance differences are not as obvious until the 2020 crash. We notice that end-to-end portfolio tracks the behavior of the nominal risk-parity portfolio until the 2020 crash. In the last year of out-of-sample period, we observe the strategies perform differently. Measuring performance over the entire out-of-sample period, Sharpe-based end-to-end risk budgeting portfolio enjoys higher risk-adjusted returns than nominal risk parity in terms of both metrics ($1.15>0.79$, $5.89>3.30$). On the other hand, return-based end-to-end portfolio provides slightly higher cumulative wealth ($1.30>1.25$) but  lower risk-adjusted performance ($0.58<0.79$) over the period 2017-2021. Notice that cumulative return as a training objective results in worse performance than Sharpe ratio. Therefore, we choose Sharpe ratio as the main training objective in the neural network for the end-to-end portfolio model. \\
\\
End-to-end Sharpe-based risk budgeting portfolio has 7.55\% annualized return whereas nominal risk parity and fix-mix strategy lost by -1.52\% and -4.34\% after July 2014 until July 2015 (Figure \ref{fig:after2013}). During that period, commodity indices suffered whereas stocks were the best performers. DBC and GLD lost by 32\% and 12\%, respectively. On the other hand, VTI and IWM had positive returns by 7\%. Even though bond indices had positive returns in this period, they had bad performance during June 2013 which caused the drop in the values of the risk budgeting portfolios. On June 19th, Ben Bernanke announced phasing out the quantitative easing program over the coming months, and investors withdrew \$23 billion from bond funds in the following week (\cite{bondcrash2013}). From Table \ref{tab:e2e-stats} we see that Sharpe-based end-to-end portfolio had less maximum drawdown value than nominal risk parity. All three bond indices have low but positive returns in the following year. End-to-end risk budgeting portfolio has more weight on municipal bonds which was the best one among three bond indices. Due to the learned dynamic risk budgets, portfolio weights are much less on corporate and treasury bonds. On the other hand, nominal risk parity have almost similar weights on three indices with heavier allocation on treasuries. Equal risk budget constraint doesn't allow to adjust the portfolio allocation based on the market circumstances, resulting in the performance difference after the crash period. On the other hand, end-to-end risk budgeting portfolio allows dynamic allocation by adjusting target risk budgets. Furthermore, it is responsive to market dynamics in commodities and equities.\\ 
\\
Similarly, in 2020, all three strategies recorded losses on March 2020 due to Covid-19 pandemic. The equal weight portfolio had 20\% drawdown whereas end-to-end and nominal portfolio values dropped by 9.5\% and 13.6\% respectively (Table \ref{tab:e2e-stats}). However, they record positive yearly returns, the end-to-end risk budgeting portfolio has the best performance with the 18\% annual return. Nominal risk parity increases by 7\% at the end of the year. Figure \ref{fig:aftercovid} presents allocation differences during and aftermath of March 2020 downturn for risk budgeting portfolios. We can see that end-to-end risk budgeting portfolio shifts majority of the allocation from municipal bonds to gold in April 2020. End-to-end portfolio's allocation shifts to commodity and equity indices according to market dynamics whereas risk parity portfolio kept more weights on bond indices due to their low volatility nature.\\
\\
We observe that end-to-end portfolio has more allocation switches than nominal risk parity. This might affect gains under transaction costs and can be mitigated by incorporating transaction penalty, but this analysis is out of the scope of this paper.

\begin{table}[hbt!]
    \centering
\begin{tabular*}{\textwidth}{c@{\extracolsep{\fill}} cccccc}
\toprule
{\textbf{Portfolios}} &   Return & Volatility &   Sharpe &     MDD & Calmar Ratio & Return/Ave.DD \\         \\
\midrule
\multicolumn{1}{c}{\textbf{2011-2016/12}}\\
e2e-sharpe     &  0.0515 &     0.0460 &  \textbf{1.1095} &  0.0501 &       1.0142 &       \textbf{ 4.5643} \\
e2e-return     & 0.0380 &     0.0466 &  0.8048 &  0.0750 &       0.5001 &        2.7017 \\
Nominal RP    &  0.0227 &     0.0359 &  0.6203 &  0.0675 &       0.3301 &        1.5307 \\
Fix-mix (1/n) &  0.0290 &     0.0689 &  0.4135 &  0.1384 &       0.2059 &        0.9327 \\
\midrule
\multicolumn{1}{c}{\textbf{2017-2021/06}}\\
e2e-sharpe    &  0.0980 &     0.0730 &  \textbf{1.1559} &  0.0949 &       0.8818 &        \textbf{5.8857} \\
e2e-return     &  0.0718 &     0.1017 &  0.5758 &  0.1542 &       0.3775 &        2.4342 \\
Nominal  RP   &  0.0609 &     0.0604 &  0.7906 &  0.1362 &       0.3484 &        3.2988 \\
Fix-mix (1/n) &  0.0943 &     0.0970 &  0.8331 &  0.2065 &       0.3894 &        3.8598 \\
\bottomrule
\end{tabular*}
    \caption{Annualized portfolio performance statistics over the in-sample (2011-2016) and out-of-sample (2017-2021/06) periods.}
    \label{tab:e2e-stats}
\end{table}


\subsection{Model-based vs. Model-free Portfolios}
\begin{figure}[hbt!]
    \centering
    \begin{subfigure}{0.9\textwidth}
    \centering
    \includegraphics[width=1.15\textwidth,center]{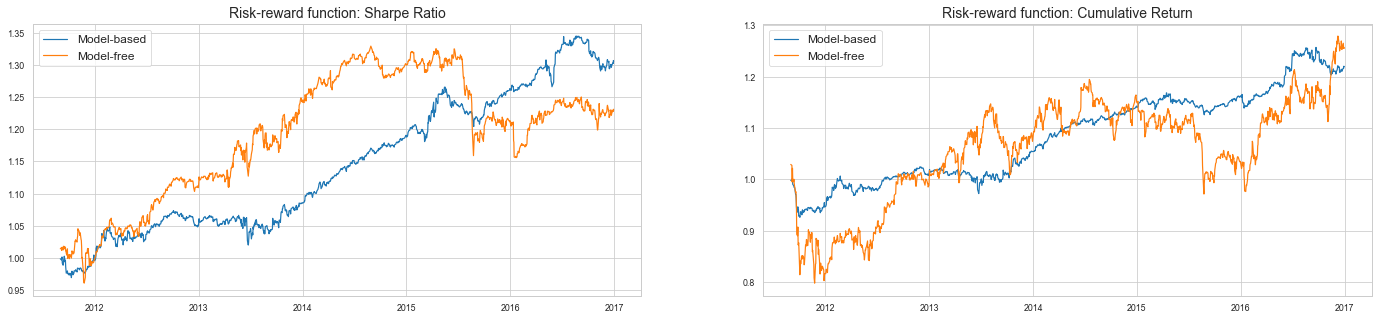}
    \caption{In sample period}
    \end{subfigure}
        \begin{subfigure}{0.9\textwidth}
    \centering
    \includegraphics[width=1.15\textwidth,center]{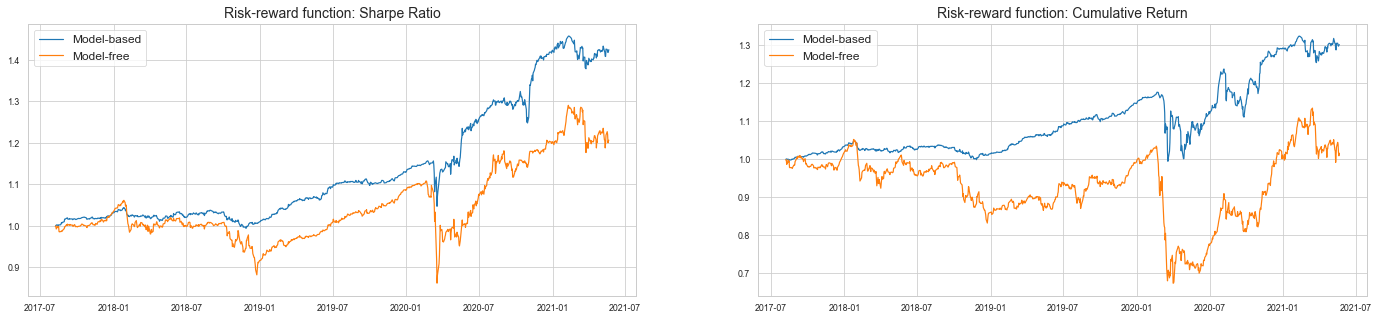}
    \caption{Out of sample period}
    \end{subfigure}
    \caption{Model-based and model-free end-to-end portfolios with seven ETFs over in-sample (2011-2016) and out-of-sample (2017-2021/06) periods. Models are trained with same hyperparameters}
    \label{fig:modelbased-vs-modelfree}
\end{figure}

In this part of the experiments, we look at differences between model-based and model-free end-to-end portfolios on the real market data. In the model-free end-to-end models, optimization layer is removed and we have a single layer fully connected neural network that learns the allocation directly from input features. Remaining parameters are kept the same in order to have a fair comparison between model-based and model-free portfolios. Similar to the conclusion in Section \ref{sec:sim-study}, model-based portfolios are more robust and stable in the real market data as well. In particular, model-free portfolio lacks any structure that guides the allocation, resulting an easy over-fitting with local structures that can be harmful when market dynamic shifts. Figure \ref{fig:modelbased-vs-modelfree} presents portfolio performances over the sample period, and quantitative metrics are available in Table \ref{tab:model-based-and-free-stats}. Sharpe-based learning results in lower drawdowns than return based learning.  In addition, performance difference between model-based and model-free portfolios is apparent with Sharpe-based learning. Specifically in out-of-sample period, model-based portfolio achieves about four times better risk-adjusted return than its model-free counterpart with Sharpe-based learning. Furthermore, model-based portfolio enjoys lower drawdowns than model-free portfolios in both time periods.

\begin{table}[]
    \centering
\begin{tabular*}{\textwidth}{c@{\extracolsep{\fill}} cccccc}
\toprule
{\textbf{Portfolios}} &   Return & Volatility &   Sharpe &     MDD & Calmar Ratio & Return/Ave.DD \\         \\
\midrule
\multicolumn{1}{c}{\textbf{2011-2016/12}}\\
\multicolumn{1}{c}{\textit{e2e-sharpe}}\\
model-based    &  0.0515 &     0.0460 &  \textbf{1.1095} &  0.0501 &       1.0142 &        \textbf{4.5643} \\
model-free  &  0.0397 &     0.0695 &  0.5640 &  0.1308 &       0.2998 &        1.2297 \\
\multicolumn{1}{c}{\textit{e2e-return}}\\
model-based  &  0.0380 &     0.0466 &  \textbf{0.8048} &  0.0750 &       0.5001 &        \textbf{2.7017} \\
model-free   &  0.0436 &     0.1431 &  0.3014 &  0.2245 &       0.1922 &        0.6190 \\
\midrule
\multicolumn{1}{c}{\textbf{2017-2021/06}}\\
\multicolumn{1}{c}{\textit{e2e-sharpe}}\\

model-based   &  0.0980 &     0.0730 &  \textbf{1.1559} &  0.0949 &       0.8818 &        \textbf{5.8857 }\\
model-free   &  0.0508 &     0.1194 &  0.3168 &  0.2226 &       0.1692 &        0.6979 \\
\multicolumn{1}{c}{\textit{e2e-return}}\\
model-based   & 0.0718 &     0.1017 &   \textbf{0.5758} &  0.1542 &       0.3775 &        \textbf{2.4342} \\
model-free   &  0.0040 &     0.1812 &  -0.0463 &  0.3587 &      -0.0218 &       -0.0633 \\
\bottomrule
\end{tabular*}
    \caption{Annualized  performance statistics of end-to-end (e2e) model-based and model-free portfolios over the in-sample (2011-2016) and out-of-sample (2017-2021/06) periods.}
    \label{tab:model-based-and-free-stats}
\end{table}
\clearpage
\section{Asset Selection in Risk Budgeting Portfolios}
\begin{figure}[hbt!]
    \centering
    \includegraphics[width=\textwidth,center]{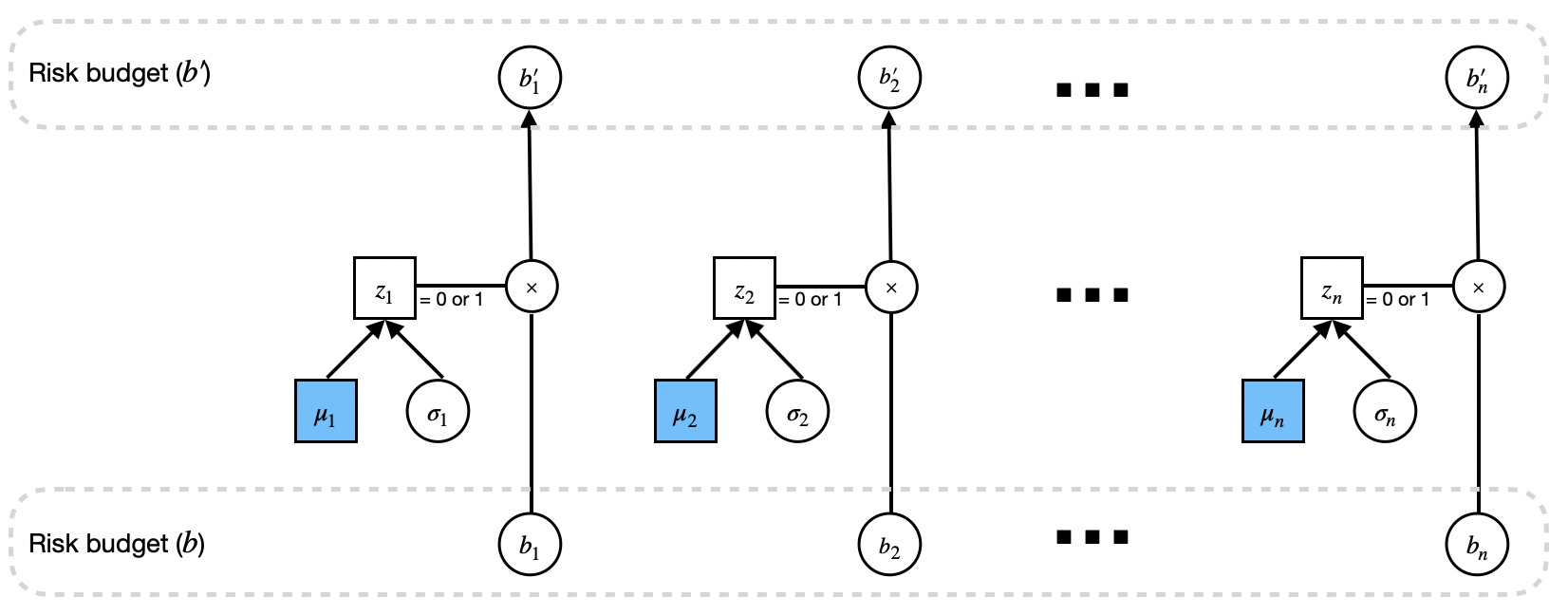}
    \caption{Stochastic gates for asset filtering.}
    \label{fig:gate}
\end{figure}

One critic of the risk-based portfolios is that although they provide robust performance most of time, they completely ignore the returns and are sensitive to underlying asset universe. The portfolio can be easily hurt by an asset with negative drift, especially if the asset also has low volatility. This drawback of risk-based portfolios could be devastating in risk-parity portfolios, due to the fact that the risk contributions of each asset is set to be the same by definition. For general risk-budgeting portfolios, the investors may mitigate the issue by wisely choosing the risk budget, but still, an ill-considered asset universe creates potential problems. With this in consideration, it is important that an investor chooses the underlying asset universe carefully. In this section, we develop a strategy incorporated in the end-to-end portfolio to provide an asset selection mechanism to boost the performance of the model-based neural networks.\\
\\
We embed a set of stochastic gates after the risk budget layer that control whether the corresponding asset risk budget passes through. The idea is that, the gates corresponding to the "bad" assets will be closed after training, and therefore the underlying asset universe on which the risk-budget portfolio based on consists only of assets with positive drifts. Let $n$ be the number of assets to select from. Following \cite{Yamada2020}, we optimize on a set of trainable parameters $\mu_1, ..., \mu_n$ that translate to the probability of the gate being active. In each training step, we introduce a small randomness $\epsilon_1, ..., \epsilon_n \sim \mathcal{N}(0,\sigma)$ so that the openness $z_d = \mu_d + \epsilon_d$ is a mean-shifted Gaussian variable. Figure \ref{fig:gate} explains the structure of the stochastic gate. In this learning task, we initialize $\mu$'s with 0.5 as a neutral starting point, and choose the randomness parameter to be $\sigma=0.1$.\\
\\
\noindent
There are two main differences between our application and the original model of \cite{Yamada2020}. First, unlike their inclusion of a penalty term in the loss function to discourage unnecessary features, we do not penalize on the activation of a gate. Recall that their purpose of stochastic gates is to select relevant features for a model with a ground-truth, and any redundant features are supposed to be filtered out. On the other hand, our purpose is merely to filter out the ``bad'' assets that are harmful to the portfolio, rather than selecting as few assets as possible. Therefore, we believe it unnecessary to penalize the inclusion of assets. Second, whereas the empirical test in \cite{Yamada2020} find the gates almost always converge to 0 or 1, the convergence in our financial application is slightly less clear. We thus apply a threshold of 0.5 in determining whether an asset should be included. With the trained parameters $\mu$ and the risk budget $b$, we apply the risk budget $b' = b \odot (\mu\geq0.5)$ to get the asset allocation, where $\odot$ is the element-wise multiplication, and $(\mu\geq0.5)$ is a vector of zeros and ones corresponding to the Boolean value.
\begin{figure}[hbt!]
    \centering
    \includegraphics[width=\textwidth,center]{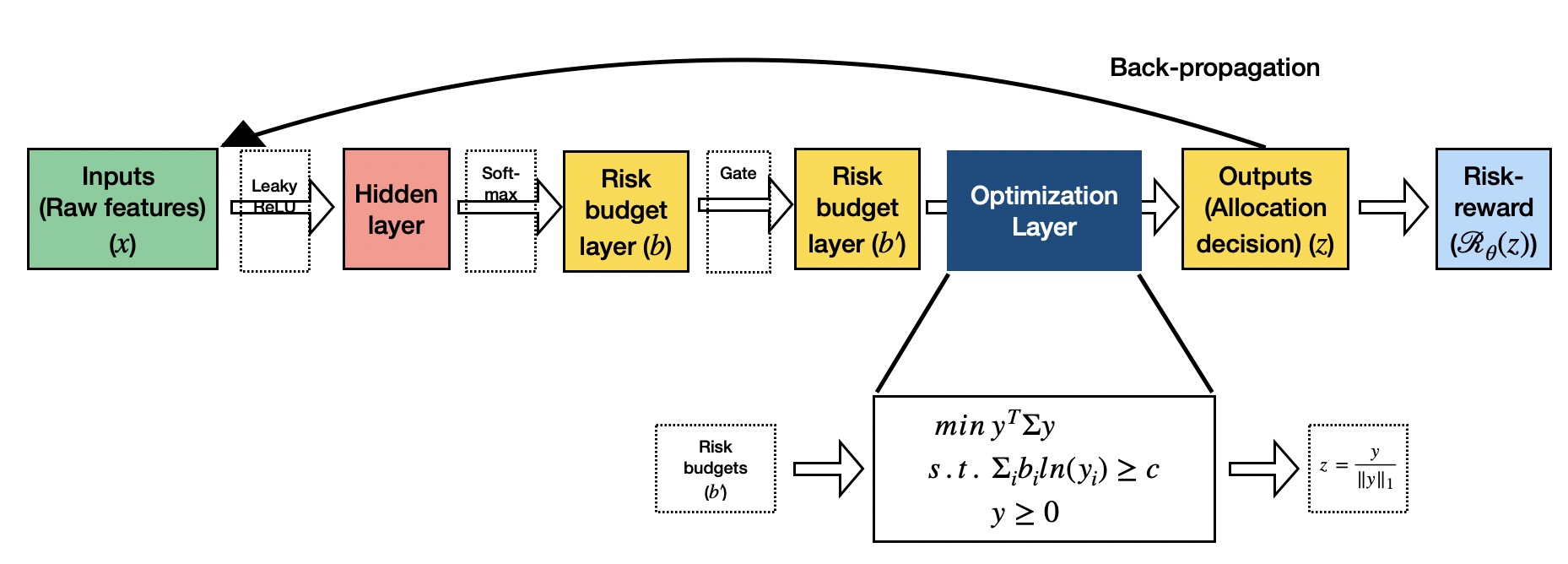}
    \caption{Computational graph of model-based approach with asset filtering gates.}
    \label{fig:computational_graph_based_gate}
\end{figure}
We apply and compare the following methods: (i) \textit{no gate}: the original model-based neural network; (ii) \textit{gate with no filter}: applying the adjusted risk budget $b'$ directly on the original covariance matrix to learn the allocation; and (iii) \textit{gate with filter}: applying the adjusted risk budget $b'$ on the covariance including only the selected assets so that all weights are given to the selected ones. 
We test the approaches on both the market data, as well as a simulated data set where we include a random asset with low return and low volatility. Such a random asset would hurt the benchmark risk-parity portfolio, but not our end-to-end strategy with asset selection procedure.

\subsection{Performance on Real Market Data}
\begin{figure}[hbt!]
    \centering
    \includegraphics[width=\textwidth,center]{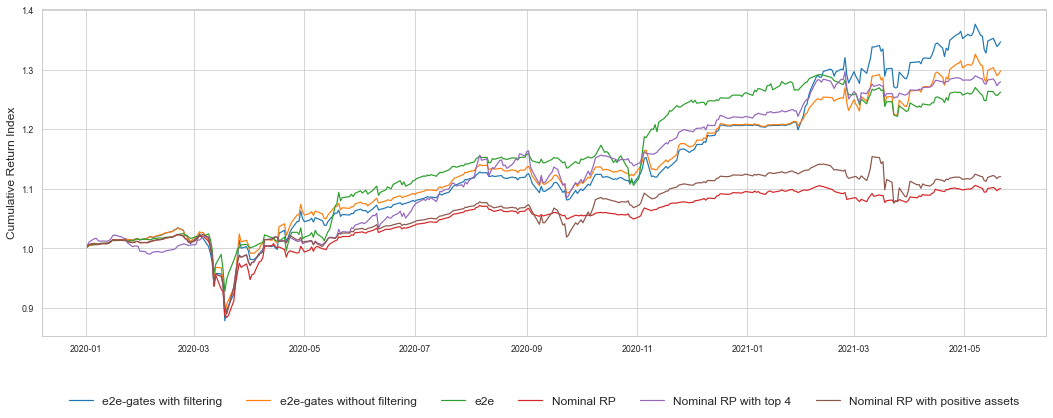}
    \caption{Out-of-sample performance of end-to-end portfolios with stochastic gates and nominal benchmarks after 2020.}
    \label{fig:portfolio-market-selection}
\end{figure}

Asset selection property in end-to-end risk budgeting portfolios is tested on the ETF investment universe. We provide three different nominal benchmarks: (1) Nominal RP: nominal risk parity, (2) Nominal RP-positive: nominal risk parity invested in assets with positive returns in the past 30-day, (3) Nominal RP-top$k$: Nominal risk parity invested in top $k$ assets\footnote{$k$ is set to be 4 in the real market data as a proxy for the top half.} in the past 30-day period. We consider a similar hyperparameter search framework presented in Section \ref{subsec:hyperparameter}. Here we have an additional hyperparameter which is the learning rate for gates ($\eta_{\mu}$). In order to be consistent, we keep the learning rate and number of training steps same ($\eta = 150$, $n=10$), and tune the learning rate of gates over the range $\eta_{\mu} = \{0.001,0.01,0.1, \dots, 100,500,1000,2000\}$. Return to average drawdown ratio provides better performance measurement for asset selection model, and we choose the optimal learning rate based on train and validation sets performance average.  \\ 
\\
We present out-of-sample results for the portfolios with and without asset selection property. Quantitative results are in Table \ref{tab:market-selection} and cumulative performance after 2020 is in Figure \ref{fig:portfolio-market-selection}. Among nominal benchmarks, the risk parity invested on top-$k$ has the best performance with a 10.5\% annualized return over 2017-2021/06. All three end-to-end portfolios beat the nominal risk parity benchmark and provide better risk-adjusted return than top-$k$ selection in terms of return to average drawdown ratio. End-to-end portfolio with stochastic gates embedding achieves a Sharpe ratio of 1.24 with the addition of the asset filtering property. In contrast, nominal risk parity and end-to-end risk budgeting portfolios have Sharpe ratios of 0.79 and 1.16, respectively. We observe that the asset filtering feature improves the portfolio performance in the risk budgeting portfolios in end-to-end and nominal models. Even with the dynamic risk budgeting option, namely e2e, we observe that small risk budgets still can lead to significant weight allocations. Ideally, we want to filter the "bad" assets out of the portfolio. Therefore, we introduce this feature with stochastic gates, and computational results justify our hypothesis.

\begin{table}[hbt!]
    \centering
\begin{tabular*}{\textwidth}{c@{\extracolsep{\fill}} cccccc}
\toprule
{\textbf{Portfolios}} &   Return & Volatility &   Sharpe &     MDD & Calmar Ratio & Return/Ave.DD \\         \\
\midrule
e2e-gate-with-filter       &  0.1233 &     0.0884 &  \textbf{1.2373} &  0.1507 &       0.7215 &        4.8034 \\
e2e-gate-no-filter       &  0.1104 &     0.0850 &  1.1380 &  0.1412 &       0.6803 &        4.8196 \\
e2e       &   0.0980 &     0.0730 &  1.1559 &  0.0949 &       0.8818 &        \textbf{5.8857} \\
Nominal RP          &   0.0609 &     0.0604 &  0.7906 &  0.1362 &       0.3484 &        3.2988 \\
Nominal RP-positive &   0.0702 &     0.0765 &  0.7448 &  0.1304 &       0.4353 &        2.1512 \\
Nominal RP-top$k$    &  0.1049 &     0.0769 &  1.1871 &  0.1292 &       0.6945 &        3.8993 \\
\bottomrule
\end{tabular*}
\caption{Annualized performance metrics over the out-of-sample period 2017-2021/06. e2e portfolio parameters: $\eta=150$, $n=10$. e2e with stochastic gates: $\eta=150$, $\eta_{\mu}=10$, $n=10$.}
\label{tab:market-selection}
\end{table}
\subsection{Performance with a Low Volatility and Low Return Asset}
\begin{figure}[hbt!]
    \centering
    \includegraphics[width=\textwidth,center]{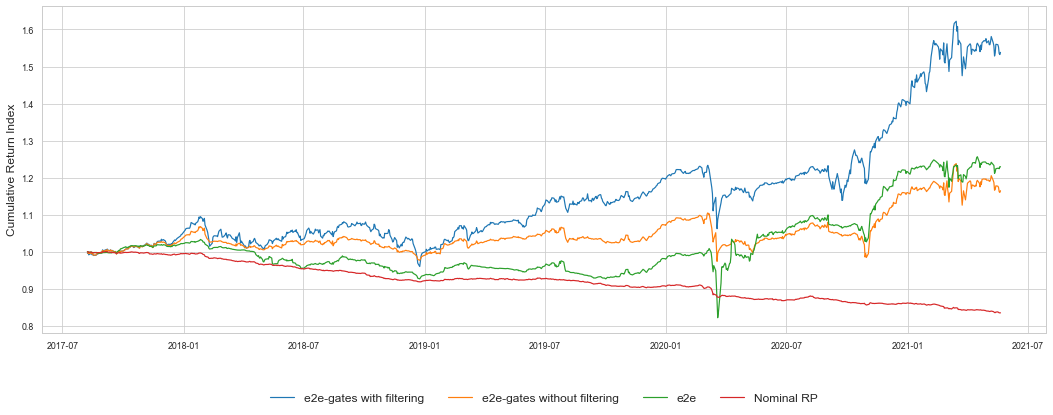}
    \caption{Out-of-sample (2017-2021/06) performance of end-to-end portfolios with stochastic gates and benchmarks.}
    \label{fig:portfolio-random-selection}
\end{figure}

Risk parity allocation structure is sensitive to asset characteristics in the portfolio, especially to low volatility assets. In order to test the effectiveness of the asset selection feature, we introduce a random asset with low volatility and return characteristics to our market portfolio. We simulate the random asset with a normal distribution with the parameters $\mu = -0.0005 $, $\sigma =0.0005 $. Table \ref{tab:random-selection} presents statistics for the portfolios with the random asset. Notice that nominal risk parity experiences the worst performance due to the tendency to allocate in the low volatility asset. On the other hand, asset filtering feature is beneficial in end-to-end risk budgeting portfolios by avoiding investment in low-volatility assets that happen to possess low returns. Performance difference between end-to-end and nominal risk parity is obvious in Figure \ref{fig:portfolio-random-selection}.\\
\\
Interestingly, both nominal filtering methods provide promising performance, with selecting the top winners improving the nominal risk parity portfolio returns significantly and producing positive returns over the out-of-sample period. 
This is not surprising because by design of this experiment, looking at past winners will guarantee the elimination of the bad-performing random asset. 
However, this is not a realistic assumption for the real market environment. We implement this example to show the benefits of the asset filtering feature in the risk budgeting portfolios, specifically in risk parity. In particular, our e2e method with filter  still yields encouraging performance in terms of both Sharpe ratio and return over average drawdown, with the highest annualized return among all tested strategies.

\begin{table}[hbt!]
    \centering
\begin{tabular*}{\textwidth}{c@{\extracolsep{\fill}} cccccc}
\toprule
{\textbf{Portfolios}} &   Return & Volatility &   Sharpe &     MDD & Calmar Ratio & Return/Ave.DD \\         \\
\midrule
e2e-gate-with-filter      &   0.1205 &     0.1194 &   0.8929 &  0.1387 &       0.7660 &        3.2742 \\
e2e-gate-no-filter         &   0.0411 &     0.0872 &   0.3234 &  0.1185 &       0.2258 &        0.6287 \\
e2e       &   0.0560 &     0.1015 &   0.4236 &  0.2046 &       0.1815 &        0.7647 \\
Nominal RP          &  -0.0465 &     0.0162 &  -3.5952 &  0.1655 &      -0.2868 &       -0.5445 \\
Nominal RP-positive &   0.0702 &     0.0765 &   0.7448 &  0.1304 &       0.4353 &        2.1512 \\
Nominal RP-top$k$     &   0.0993 &     0.0762 &   1.1241 &  0.1292 &       0.6525 &        3.5327 \\
\bottomrule
\end{tabular*}
\caption{Annualized performance metrics over the out-of-sample period 2017-2021/06. e2e portfolio parameters: $\eta=500$, $n=5$. e2e with stochastic gates: $\eta=750$, $\eta_{\mu}=750$, $n=10$.}
\label{tab:random-selection}
\end{table}
\section{Conclusion} \label{sec:conclusion}
This paper adopts an end-to-end neural network approach to tackle the portfolio allocation problem with a model-based and model-free setup. The model-based approach learns the target risk budgets in the portfolio and allocates them according to the risk-budgeting strategy, and the model-free approach learns the allocation directly from the input features. We observe that the model-based end-to-end portfolio provides robust and satisfying performance on the real market data as well as in the simulation study. In particular, it outperforms nominal risk-parity and equal weights (1/N) benchmark strategies on market data regarding Sharpe ratio and return over average drawdown. The allocation of model-based end-to-end portfolio tracks the nominal risk-parity under normal market conditions and presents advantages under abnormal events such as bond crash and Covid-crash by effectively learning the recent dynamics. We introduce an asset selection feature that lets us filter out the unprofitable assets from the investment universe. Not only it protects the risk budgeting portfolio against the under performing low volatility assets, but it also boosts the performance in the real market data environment. 
\subsection{Future Work}
There are numerous next steps that we want to point out for future research to enhance end-to-end portfolio optimization with risk budgeting layer. Here we choose to use sample estimate of covariance matrix since it has been shown that risk budgeting optimization problem robust to parameter estimates. However, it can be treated as a parameter in the optimization layer and determined in the learning process. In this paper, we treated multi-period problem by solving single-period optimization problem in every step and assumed no transaction cost environment. On the other hand, with a multi-period problem we can address transaction costs and regulate portfolio turnover rate. The model predictive control approach in \cite{uysalli2021} can be implemented to construct multi-period risk budgeting portfolios. Furthermore, computational results can be extended to equity portfolio to test the performance in difference asset universes. This end-to-end framework can be constructed with a different portfolio optimization model of choice.

\clearpage
\newpage
\bibliographystyle{abbrvnat}
\bibliography{main}
\newpage
\end{document}